\documentclass[12pt,preprint]{aastex}
\usepackage[nonamelimits]{amsmath}
\usepackage{graphicx}
\usepackage[onecolumn]{emulateapj5}

\def \oneight{RX J1856.5-3754}
\def \zeroseven{RX J0720.4-3125}

\newcommand{\bc}{{\cal {B}}}
\newcommand{\ac}{{\cal {A}}}
\newcommand{\dc}{{\cal {D}}}
\newcommand{\cc}{{\cal {C}}}
\begin{document}


\title{Bare Quark Stars or Naked Neutron Stars ? \\The Case of \oneight}

\author{Roberto Turolla\altaffilmark{1}, Silvia Zane\altaffilmark{2} and
Jeremy J. Drake\altaffilmark{3}}

\altaffiltext{1}{Department of Physics, University of
Padova, via Marzolo 8, 35131 Padova, Italy; turolla@pd.infn.it}
\altaffiltext{2}{Mullard Space Science Laboratory, University College London,
Holmbury St. Mary, Dorking, Surrey, RH5 6NT, UK; sz@mssl.ucl.ac.uk}
\altaffiltext{3}{Smithsonian Astrophysical Observatory, MS 3, 60 Garden
Street, Cambridge, MA 02138, USA; jdrake@head-cfa.harvard.edu}


\begin{abstract}

In a cool neutron star ($T \lesssim 10^6$~K) endowed with a rather
high magnetic field ($B \gtrsim 10^{13}$~G), a phase transition
may occur in the outermost layers. As a consequence the neutron
star becomes ``bare'', i.e. no gaseous atmosphere sits on the top
of the crust. The surface of a cooling, bare neutron star does not 
necessarily emit a blackbody spectrum because the emissivity is 
strongly suppressed at energies below the electron plasma frequency,
$\omega_p$. Since $\omega_p\approx 1$~keV under
the conditions typical of the dense electron gas in the
condensate, the emission from a $T\sim 100$ eV bare neutron star
will be substantially depressed with respect to that of a perfect
Planckian radiator at most energies. Here we present a detailed
analysis of the emission properties of a bare neutron star. In
particular, we derive the surface emissivity for a Fe composition
in a range of magnetic fields and temperatures representative of
cooling isolated neutron stars, like \oneight. We find that the
emitted spectrum is strongly dependent on the electron
conductivity in the solid surface layers. In the cold electron
gas approximation (no electron-lattice interactions), the
spectrum turns out to be a featureless depressed blackbody in the
0.1--2~keV band with a steeper low-energy distribution. When
damping effects due to collisions between electrons and the ion
lattice (mainly due to electron-phonon interactions) are
accounted for, the spectrum is more depressed
at low energies and spectral features may be present,
depending on the magnetic field strength. Details of
the emitted spectrum are found, however, to be strongly dependent
of the assumed treatment of the transition from the external vacuum 
to the metallic surface. The implications of out results to \oneight \ 
and other isolated neutron stars are discussed.

\end{abstract}


\section{Introduction}
\label{intro}

More than 20 X-ray sources currently associated with isolated
neutron stars (INSs) show evidence for a thermal component in their
spectrum, in many cases superimposed on a power-law high-energy tail.
In the commonly accepted picture the hard tail is produced
by non-thermal processes in the stellar
magnetosphere. On the contrary, the
thermal component originates at the surface, while the
star cools down and internal energy is progressively radiated away.
Since these objects are not complicated by strong accretion
signatures, detailed observations of
their thermal component provide a powerful tool to investigate directly
the properties of the neutron star (NS). If the thermal emission originates
in the NS atmosphere, the detection or absence of spectral lines and edges may
constrain the chemical composition and/or magnetic field through a
comparison with computed models (see e.g. \citealt{shi92}; \citealt{rrm97};
\citealt{pons2002}). Furthermore, empirical insights may be derived for the
NSs mass, radius and equation of state (e.g. \citealt{lapra2001}).

The family of thermally emitting INSs includes seven peculiar
objects serendipitously discovered in {\it ROSAT\/} PSPC pointings (see
e.g. \citealt{t2000} and \citealt{m2001} for reviews;
\citealt{zam2001}). These
sources (hereafter referred to as {\it ROSAT\/} INSs) are characterized
by remarkably similar properties among which are: soft, thermal
spectrum with $kT\sim 100$~eV; low X-ray luminosity,
$L_X\approx 10^{30}-10^{31} {\rm erg\,s}^{-1}$, and low column density,
$N_H\sim 10^{20} \ {\rm cm}^{-2}$; no association with a
supernova remnant; pulsations in the 5--20 s range (detected in
four sources so far). Until recently the spectral properties of
the seven {\it ROSAT\/} INSs were not known in detail. PSPC observations
provided evidence that
a blackbody spectrum gives a satisfactory description of the
data in all cases. However, the scanty statistics prevented definite
conclusions being drawn about the X-ray spectral energy
distribution (SED), and did not allow for the detection of
spectral features. While this situation has not yet improved for
the fainter sources, the two  brightest {\it ROSAT\/} INSs, \oneight \ and
\zeroseven, have been the target of deep observations
with {\it Chandra\/} and {\it XMM-Newton\/}.
\zeroseven \ was
observed with {\it XMM-Newton\/} for the first time on May~2000,
in a 62.5~ks pointing (\citealt{pae2001}; \citealt{cro2001}). The
EPIC-PN spectrum is well-represented by a blackbody and no
spectral features have been detected, apart from variations in the column
density with pulse phase that may be explained in terms of
energy-dependent beaming effects or cyclotron absorption
(\citealt{pae2001}; \citealt{cro2001}; \citealt{hab2003}).
The X-ray flux shows a modulation with a period of 8.31~s and a pulsed
fraction of $\sim 15\%$ (\citealt{ha97}; \citealt{cro2001}).
Using {\it ROSAT\/} and {\it XMM-Newton\/} data, \citet{silvia2002} were
able to derive the period derivative
($\dot P \sim 5\times 10^{-14}\, {\rm ss}^{-1}$) which,
when interpreted in terms of
magneto-dipolar braking, implies a surface magnetic field of
$\sim 2\times 10^{13}$ G. The case of \oneight \ is even more
striking. A 50 ks {\it Chandra\/} LETGS observation has
convincingly shown that \oneight \ appears to have a featureless
X-ray continuum, for which a simple blackbody yields a better fit
than more sophisticated atmosphere models \citep{bur2001}.
Analysis of a very long ($\sim 500$ ks)
{\it Chandra\/} observation obtained in October 2001
further reinforces this conclusion and rules out the presence of
strong spectral features \citep{dra2002}. However, the presence of
broadband departures from a pure blackbody spectrum have been claimed
by Burwitz et al. (2003) in
both {\it XMM-Newton} and {\it Chandra} data.
No X-ray pulsations have been detected and the upper limit on the pulsed
fraction is now down to
$\lesssim 1.3\, \%$ (\citealt{rgs2002}; \citealt{dra2002}; \citealt{bur2003}).
\oneight \ was the first source in this class for which an
optical counterpart has been found (\citealt{wm97}; \citealt{vkk2001a}).
When used in conjunction with {\it Chandra\/} data, the recently
measured star distance (about 120~pc; \citealt{kka2001};
\citealt{wl2002}) yields a radiation radius of only $\sim 5$-6 km
\citep{dra2002}. Very recently \cite{hab2003} reported the
discovery of spectral feature(s) in the EPIC/RGS data of RBS 1223.
The feature, seemingly present also in {\it Chandra} data, is a deficit
of counts with respect to the best-fitting blackbody at an energy of $\sim
0.2-0.3$~keV, and is quite broad. Its nature is still
uncertain but it could be a proton cyclotron
absorption line in a magnetic field $\sim 4\times 10^{13}$~G, as
discussed by \cite{hab2003}.

The small apparent radius and the blackbody X-ray spectrum led to
the intriguing suggestion that \oneight \ might be a strange
quark star (\citealt{dra2002}; \citealt{xu2002}). One of the
motivations for such a claim is that bare quark stars, i.e. those
not covered by a layer of hadronic matter, do not have an
atmosphere and would {\it presumably} emit a pure blackbody
spectrum (as suggested for instance by \citealt{xu2002}).
However, further investigations are definitely needed to assess
the spectral properties of these objects. While a quark star may
be a conceivable option (see however \citealt{thoma2003}), present
observations of \oneight \ do not necessarily demand this
solution and more conventional scenarios involving a neutron star
are certainly possible. Neutron star models based on a
two-temperature surface distribution can account for both the
X-ray and optical emission of \oneight, giving at the same time
acceptable values for the stellar radius (\citealt{pons2002};
\citealt{wl2002}; \citealt{braro2002}). However, the problem of
producing a featureless spectrum from a NS has not been
consistently solved yet, although possible ways to suppress the
spectral features which are expected in optically thick
atmospheric models have been outlined (see e.g.
\citealt{braro2002}).

In this paper, we consider an alternative explanation for the
peculiar X-ray spectrum of \oneight. In the 70's it was commonly
accepted that radiation emitted by NSs came directly from their
solid surface (e.g. \citealt{bri80} and references therein).
Later, the role of the thin gaseous layer which covers the star
crust in shaping the emergent radiation spectrum was appreciated
and model atmospheres became the standard tool for interpreting
the observed emission from isolated NSs. However, highly
magnetized neutron stars may be left without an atmosphere if
they are cool enough. The reason for this is the onset of a phase
transition that turns the gaseous atmosphere into a solid when
the surface temperature drops below a critical temperature
$T_{crit}$, which, for a given chemical composition, depends on
the stellar magnetic field (see \citealt{laisalp97} and
\citealt{lai2001} for a recent review). The determination of
$T_{crit}$ is still uncertain and, in particular, only preliminary
calculations are presently available for heavy element (such as
Fe) surface compositions  (\citealt{lai2001} and references
therein). In \S \ref{bare} we show that, given the large
uncertainties on the conditions for Fe condensation, it is
possible that \oneight \ and (marginally) \zeroseven \ have
surface temperatures below the critical value and may be then
``naked'' or ``bare'' neutron stars.

The idea that \oneight \ might be a solid surface NS was
suggested earlier by \cite{bur2001} and \cite{za03} (see also
\citealt{bur2003}). As first discussed by \cite{trule78}, a
severe reduction in the NS surface emissivity occurs at energies
below the electron plasma frequency. Under the conditions for
which a {\it ROSAT} INS is bare ($T\lesssim 100$~eV, $B\gtrsim
10^{13}$~G), the plasma frequency in the surface layers
corresponds to energies $\gtrsim 1$~keV (see \S~\ref{bare}), so 
the NS is expected to
radiate less efficiently than a blackbody emitter at soft X-ray
energies and below. This is of great potential importance, since
it might help reconcile the observed radiation radius with current
theoretical predictions of NSs radii. Also, it may hold the key
for explaining the featureless blackbody spectrum observed in
some of these sources. In \S~\ref{emission} we address in detail
the question of the nature of the surface emissivity of a bare NS with
a pure Fe composition. The method we use is similar to that
employed by \cite{bri80}, who was the first to investigate this
issue in connection with X-ray pulsars. We found that the
emissivity, and hence the shape of the emitted spectrum, depends
crucially on the conductivity of the star crust. 
In \S~\ref{cold} we analyze a simple (albeit
unrealistic) model in which only the
contribution of a cold electron plasma to the dielectric
tensor in the star interior is accounted for. Results for this case are
qualitatively similar to those of \cite{bri80}. Proper account
for the damping produced by interactions of (degenerate)
electrons with the ion lattice (mainly through electron-phonon
collisions, e.g. \citealt{pot99}), however, introduces
qualitative changes to the above picture, as is discussed in
\S~\ref{damp}.  The relevance of our model to \oneight \ is
finally discussed in \S~\ref{discuss}.


\section{Bare Neutron Stars}
\label{bare}

In this section, we explore the possibility that some of the
cooler isolated neutron stars ($T\lesssim 100$ eV) are left
without an atmosphere by a phase transition in the surface layers
at large magnetic fields ($B\gtrsim 10^{13}$ G). We point out
that, although this is
unlikely for a light element (H, He) composition, it might be the
case for heavy elements (such as Fe), at least for some sources
notably including \oneight. If
indeed some INSs are bare, the question of the nature of their emitted
spectrum arises. In a neutron star with metallic surface layers
(here and in the following $Z$ and $A$ denote the atomic number and weight
of the
constituent element) the density at zero pressure is given by
(e.g. \citealt{lai2001})

\begin{equation}
\label{dens}
\rho_s \approx 560 A Z^{-3/5} B_{12}^{6/5} \, {\rm g \, cm}^{-3}\, ,
\end{equation}
where $B_{12}=B/10^{12}$~G.
The electron plasma frequency is then
\begin{equation}\label{omegap}
\hbar\omega_p=\hbar\sqrt{\frac{4\pi
e^2n_e}{m_e}}\approx 0.7\, Z^{1/5}B_{12}^{3/5}
\left(\frac{\rho}{\rho_s}\right)^{1/2}\, {\rm keV},
\end{equation}
where $n_e$ is the electron density. In the following we take as a reference
value for the plasma frequency that given by
eq.~(\ref{omegap}) with $\rho = \rho_s$ and $Z=26$; $\omega_{p,0}$ represents
then the plasma frequency in a pure iron medium with density $\rho_s$
assuming that all electrons are in the conduction zone (see \S \ref{cold}
for a further discussion).

Cool NSs ($T\lesssim 100$ eV) emit most of
their thermal radiation below the plasma frequency and
substantial deviations from a pure blackbody spectrum are
expected as a result of the large absorption at $\omega\lesssim \omega_p$.
Spectral features should also appear around the
electron cyclotron frequency at
\begin{equation}
\hbar\omega_B= \frac{eB}{m_ec}\simeq 11.6\, B_{12}\, {\rm
keV}\, ,
\end{equation}
but, since we focus on field values $B\gtrsim 10^{13}$ G,  these fall well
outside the X-ray range accessible to the {\it Chandra\/} LETGS and {\it
XMM-Newton\/} EPIC-PN, and are of no immediate interest.

The properties of atoms and condensed matter are qualitatively
changed by magnetic effects when $b = B/B_0 \gg 1$, $B_0 =
m_ee^3c/\hbar^3 \simeq 2.35 \times 10^9$~G. Theoretical research
on matter in superstrong fields started over 40 years ago and
although many uncertainties still remain, much progress has been
made, especially for H and He compositions (see \citealt{lai2001}
and references therein). In particular, when $b \gg 1$, electrons
are strongly confined in the direction perpendicular to the
magnetic field and atoms attain a cylindrical shape. Moreover, it
is possible for these elongated atoms to form molecular chains by
covalent bonding along the field direction. Interactions between
the linear chains can then lead to the formation of
three-dimensional condensates. As discussed by \citet{laisalp97}
and \citet{lai2001}, in the case of hydrogen the infinite linear
chains (and metallic hydrogen) are certainly bound, favoring the
possibility of condensation for sufficiently low temperatures
and/or strong magnetic fields. The critical temperature below
which phase separation between condensed H and vapor occurs is
\begin{equation}
\label{thyd}
T^H_{crit} \approx 0.1 Q_\infty
\end{equation}
with
\begin{equation}
Q_\infty \approx 194.1B_{12}^{0.37} - 4.4 \left ( \ln B_{12} -
6.05 \right )^2  -
\hbar \omega_{p,p}  - \frac{\hbar}{2} \left (\omega_{B,p}^2
 +  \omega_{p,p}^2 \right )^{1/2} + \frac{1}{2} \hbar
\omega_{B,p}\,  {\rm eV}\, .
\end{equation}
where $\omega_{p,p}$ and $\omega_{B,p}$ are the proton plasma and
cyclotron frequencies.

For heavier elements (such as Fe), the lattice structure and the cohesive
properties of the condensed state are very uncertain and are different from
those of H and He. For instance, unless the
field is extremely high ($B_{12} \gg 100$), it is likely that the linear
chains are unbound for $Z \gtrsim  6$. More recent computations of
the cohesive energy $Q_s$ of the 3D condensate showed that $Q_s$ is only a
tiny fraction ($\sim 0.5$\%)
of the atomic binding energy, correcting earlier overestimates
(\citealt{jo86}, see also \citealt{nkl87}):

\begin{equation}
Q_s \lesssim 0.05 | E_{atom} | \sim Z^{9/5} B_{12}^{2/5} \,  {\rm eV}
\quad {\rm for} \, Z  \gtrsim  10 \, .
\end{equation}
On the other hand (see again \citealt{lai2001}), even such
a weak cohesion of the Fe condensate can give rise to a phase transition
at sufficiently low $T$. The critical temperature at which phase separation
occurs can be estimated by equating the ion density of the condensed
phase near zero pressure [eq. (\ref{dens})] to the gas density in the vapor
\citep{lai2001}
\begin{equation}
\rho_g\approx 390 A^{5/2}T^{5/2}\exp{(-Q_s/T)}\, {\rm g \, cm}^{-3}\, .
\end{equation}
This gives
\begin{equation}
\label{tfe}
T_{crit}^{Fe} \lesssim 0.1 Q_s \approx 27 B_{12}^{2/5} \, {\rm eV}\, .
\end{equation}
It should be noted that, although representing the more recent
available estimates, these expressions for heavier elements are
still quite crude: all models are approximate near zero pressure
and the structure itself of the lattice is very uncertain. For
our purposes, they should be regarded as being typically accurate
to an order of magnitude. Also, the vapor density becomes much less than
the condensation density and a phase transition is unavoidable
only when the temperature drops below $\sim T_{crit}/2$ (see
\citealt{lai2001}).

The critical condensation temperatures for H and Fe are plotted
as a function of $B$ in Figure~\ref{tcrit}. The filled circles
show the position in the $B$-$T$ plane of the coolest ($T\lesssim
100$ eV), thermally emitting INSs for which an estimate of the
magnetic field is available (see table \ref{tableins}). We have
also included \oneight \ in Figure~\ref{tcrit}; its position is
indicated by a horizontal line since its magnetic field is not
presently known. In order to obtain the local surface
temperature, i.e. the quantity reported in Figure~\ref{tcrit}, a
gravitational red-shift correction was applied to the values
listed in table \ref{tableins}, according to the expression
$T_{surf}=(1+z)T_{bb}$ where $(1+z)^{-1}= \sqrt{1-2GM/c^2R}\simeq
0.8$ ($M,R$ are the star mass and radius). Here $T_{bb}$ is the
color temperature, as derived from the blackbody fit.

It is apparent from Figure~\ref{tcrit} that all INSs have a
temperature well in excess of the H critical temperature: if
surface layers are H-dominated, the presence of a gaseous
atmosphere is unavoidable.  On the other hand, if INSs have not
accreted much gas, one might expect to detect thermal emission
directly from the iron surface layers.  If this is the case, the
outermost layers of \oneight \ (depending on the magnetic field),
and possibly \zeroseven, might be in form of hot condensed
matter, in which case the usual radiative transfer computations
do not apply.


\section{The Surface Emissivity}
\label{emission}

The emission properties of the neutron stars surface have been
first analyzed by \cite{trule78} and in some more detail by
Brinkmann (1980, hereafter B80). Both of these works were aimed to
X-ray pulsars, where the surface temperature is a few keV's, and
treated the medium inside the star as a cold electron plasma,
neglecting all possible effects due to electron degeneracy and
ion lattice (primarily through electron-phonon interactions). The
(constant) damping frequency which appears in B80 calculations is
mainly used to smear the resonance at the cyclotron frequency.
Moreover, birefringence in the magnetized vacuum outside the star
was not accounted for. In this section we derive the NS surface
emissivity following an approach similar to that discussed in
B80. To better illustrate the importance of electron-phonon
interactions, we first consider a pure cold electron plasma,
repeating Brinkmann's calculation for the parameter values
appropriate to cold isolated NSs (\S\ref{cold}). A complete
treatment which includes the polarization properties of
magnetized vacuum is presented in Appendix \ref{app1}. Since, as
we show there, this more general approach is quite cumbersome and
only gives tiny differences with respect to the simpler one based
on unpolarized radiation, the latter is used below. In
\S\ref{damp} we analyze the more realistic case in which the
damping of electromagnetic waves produced by the presence of the
ion lattice is included.

\subsection{The Cold Plasma Case}
\label{cold}

We start considering the medium inside the star as a cold electron
plasma and neglect the damping of free electrons due to
collisions. We introduce a cartesian frame as in B80 (see his
figures 1 and 2) with the $z$-axis parallel to the surface
normal. The direction of the incident wave vector $\mathbf{k}$ is
specified by the angle of incidence $i$ and the azimuth $\beta$.
The magnetic field direction ${ \bf b} \equiv \mathbf{B} /B$ is
at an angle $\alpha$ with respect to the $z$-axis and
$\mathbf{b}$ lies in the $x-z$ plane. Given a star surface element
$dA=2\pi R^2\sin\theta\, d\theta$ at magnetic co-latitude
$\theta$, we first compute the total reflectivity $\rho_\omega$ of
the surface for incident unpolarized radiation. Then, since the
absorption coefficient is simply $\alpha_\omega=1-\rho_\omega$,
Kirchhoff's law yields the emissivity $j_\omega = \alpha_\omega
B_\omega(T)$, where $T$ is the temperature of the emitting
element. In general, $\rho_\omega$ depends on the direction of
the refracted ray (see below). Therefore, the monochromatic flux
$f_\omega$ emitted by the surface element must be computed by
integrating over all incident directions,
\begin{equation}\label{df}
f_\omega= \int_0^{2\pi}\int_{-\pi/2}^{\pi/2}
j_\omega(i,\beta, \theta)\sin i\, di \,
d\beta\, .
\end{equation}
The flux emitted by the entire surface is  given by \footnote{
Viewing angle effects have been neglected in evaluating the flux
from eq.~(\ref{ftot}).}
\begin{equation}\label{ftot}
F_\omega=\frac{1}{4\pi R^2}\int_{sphere}f_\omega\, dA =
\frac{1}{2}\int_0^{\pi} \sin \theta\, d\theta
\int_0^{2\pi}\int_{-\pi/2}^{\pi/2}j_\omega(i,\beta,
\theta)\sin i\, di \, d\beta\, .
\end{equation}

At the surface, an incident electromagnetic wave,
described by its electric field ${\bf E}$ and wave vector ${\bf k}$, is
partly reflected (${\bf E}'', {\bf k}''$) and partly refracted.
Due to the birefringence of the medium, the refracted wave is the sum of
an ordinary (${\bf E}'_1, {\bf k}'_1$) and an extraordinary (${\bf
E}'_2, {\bf k}'_2$) mode. In order to compute the reflectivity, we need
to solve the dispersion relation and to compute the refractive index
$n$ for the two modes of propagation.
In our frame, the dielectric tensor for a cold
electron plasma is given by

\begin{equation}\label{maxw}
\epsilon_{ij}=
\left(\begin{array}{ccc}
  S\cos^2\alpha+P\sin^2\alpha & -iD\cos\alpha & \sin\alpha\cos\alpha(P-S)
\cr
  iD\cos\alpha                &  S            & -iD\sin\alpha
\cr
   \sin\alpha\cos\alpha(P-S)  & iD\sin\alpha  &
P\cos^2\alpha+S\sin^2\alpha
\end{array}\right)
\end{equation}
with
\begin{equation}
\label{rl}
\left(\begin{array}{c}R \cr L \end{array}\right)
= 1-\frac{\omega_p^2}{\omega^2}\frac{\omega}{\omega\mp\omega_B}
\, ,
\end{equation}
\begin{equation}
\label{p}
P =1-\frac{\omega_p^2}{\omega^2}
\, ,
\end{equation}
\begin{equation}
\label{sd}
\left(\begin{array}{c}S \cr D \end{array}
              \right) = (R\pm L)/2 \, .
\end{equation}

By introducing the
Maxwell tensor
$\lambda_{ij}= k'_ik'_j-
k^{\prime 2}\delta_{ij}+(\omega^2/c^2)\epsilon_{ij}$,
 where $k'_i$ are the cartesian
components of ${\bf k'}$ and $k^{\prime 2}\equiv k'_ik'_i$,
the dispersion relation is obtained by
imposing $\vert\lambda_{ij}\vert=0$. For our purposes it is
convenient to write the resulting expression in terms of
angle of incidence $i$, and the
(complex) refractive index $n=k' c/\omega$. By using an
expression formally analogous to Snell's law $n = \sin i
/\sin \Theta$ (where now $\Theta$ is a complex quantity which
replaces the angle of refraction while $i$ is real; see e.g.
\citealt{marion}), it is

\begin{eqnarray}\label{disp}
& n^4(P+v\sin^2\alpha)+n^2(gv-2PS+u\sin^2\alpha)+PRL+gu =\cr
&\sin i\sin(2\alpha)\cos\beta(n^2-\sin^2i)^{1/2}(u+n^2v)\, ;
\end{eqnarray}
in the previous expression $v=S-P$, $u=PS-RL$ and
$g=\sin^2i[1-\sin^2\alpha(1+\cos^2\beta)]$. Squaring
eq.~(\ref{disp}) gives a fourth order polynomial equation in
$n^2$ which can be solved analytically. Clearly only two out of
four solutions satisfy the original dispersion relation and
represent the refractive indices for the two propagation modes in
the magnetized plasma, $n_m$, $m=1 \, ,2$. As noted by B80, the
only practical way of finding the two meaningful roots  is to
substitute them back into eq.~(\ref{disp}) and check numerically
the residual.  This, however, turned out to be troublesome for
some values of the parameters, as we discuss later on. For $i=0$,
$\alpha =0$ or $\pi/2$, $\beta=\pi/2$ or $3\pi/2$, the right hand
side of eq.~(\ref{disp}) vanishes and the dispersion relation
reduces to a quadratic equation in $n^2$ which is then solved
instead of the quartic.

Once the refractive indices are known,
we can solve the wave equation for the two refracted modes
$\lambda_{ij}(n_m)E'_{m,j}=0$, where $E'_{m,j}$ are the
cartesian components of ${\bf E}'_m$, obtaining the
two ratios $E'_{m,x}/E'_{m,z}$ and $E'_{m,y}/E'_{m,z}$. We
performed the calculation (double-checked with the aid of an algebraic
manipulator), obtaining
\begin{eqnarray}\label{aandb}
\frac{E'_{m,x}}{E'_{m,z}} \equiv a_m & = &
\left[ - n_m^2 \sin^2 i \sin \beta \cos \beta -i D \sin^2 i \cos \alpha
+ i D \cos \beta \sin \alpha \sin i \sqrt{n_m^2 - \sin^2 i}
\right.\nonumber\\
& - & \left.  \sin
\beta \left (P-S \right ) \sin \alpha \cos \alpha \sin i \sqrt{n_m^2 -
\sin^2 i}\right.\nonumber\\
&+& \left.\sin^2 i \sin \beta \cos \beta \left (P \cos^2 \alpha + S \sin^2
\alpha \right ) + i D \cos \alpha P\right ]\nonumber\\
& \times &\left [- n_m^2 \sin i \sqrt{n_m^2 - \sin^2 i} \sin \beta + i D
\sin
\alpha n_m^2 - iD \sin \alpha \sin^2 i \cos^2 \beta\right.
\nonumber\\
&- & \left.i D \cos \alpha \cos \beta \sin i \sqrt{n_m^2 - \sin^2
i}
+ \sin i\sqrt{n_m^2 - \sin^2 i} \left [ \sin \beta S
\right.\right.\nonumber\\
&+& \left.\left.
\sin \beta \sin^2 \alpha\left (P-S \right )\right] -\left(P-S \right)
\sin \alpha \cos \alpha \sin^2 i \sin \beta \cos
\beta -iD \sin \alpha P\right]^{-1} \\
\frac{E'_{m,y}}{E'_{m,z}}  \equiv  b_m  & = &  \left[a_m\left(\sin^2i
\sin\beta\cos\beta-iD\cos\alpha\right)+\sin\beta\sin i\sqrt{n_m^2-\sin^2
i}+
iD\sin\alpha\right]\nonumber\\
&\times & \left(\sin^2\beta\sin^2 i-n_m^2+S\right)^{-1}\nonumber\,
.
\end{eqnarray}
While the previous expression for $b_m$ agrees with that given in
B80, our result for $a_m$ is different and we were unable to
recover his expression.

The ratios in eq.~(\ref{aandb}) are then inserted into the
Fresnel equations which fix the boundary conditions at the
interface between the two media (see \citealt{ja75} and
eqs.~[17]-[18] in B80). This allows the derivation of the
components of the electric field of the reflected wave parallel
and perpendicular to the plane of incidence ($E''_\parallel$ and
$E''_\perp$) in terms of the same components of the incident wave
($E_\parallel$ and $E_\perp$). We re-derived the expressions
given in B80 and found
\begin{eqnarray}\label{erefl}
E''_\parallel & = &
\dc^{-1} \left [ \frac{A_+B_- - A_-B_+}{\bc_1\bc_2 \left( 1+w_2
\right)\left( 1+w_1 \right) }  E_\perp +
\left ( A_- - B_- \right) E_\parallel \right ]\cr
E''_\perp & = &
 \dc^{-1} \left [ \left ( \frac{ 1-w_1}{1+w_1} A_+ -
\frac{ 1-w_2}{1+w_2} B_+ \right ) E_\perp + 2 \bc_1 \bc_2 \left(w_1 - w_2
\right) E_\parallel \right ]
\end{eqnarray}
where
\begin{eqnarray}
A_{\pm} & = & \bc_1 \left (1+w_1 \right) \left [ \ac_2 \left( w_2 \cos i
\pm
\frac{1}{\cos i } \right ) + \sin i \right ]\, ,\cr
B_{\pm} & = & \bc_2 \left (1+w_2 \right ) \left [ \ac_1 \left( w_1 \cos i
\pm
\frac{1}{\cos i } \right ) + \sin i \right ]\, ,
\end{eqnarray}
$w_m=\sqrt{n_m^2 - \sin^2 i}/\cos i$, $\ac_m = b_m\sin\beta-a_m\cos\beta$,
$\bc_m = b_m\cos\beta+a_m\sin\beta$ and $\dc = A_+-B_+$.
These expressions were double-checked with the aid of an algebraic
manipulator and differ again from those in B80; we also note that
our definition of  $B_{\pm}$ is different from that of B80.

The reflection coefficient for unpolarized radiation can be
expressed as the combination of the reflectivity of parallel and
perpendicularly polarized incident waves, $\rho_\omega =
(\rho_{\parallel,\omega} + \rho_{\perp,\omega})/2$. Since the
reflectivity is defined as the ratio of the reflected to the
incident wave amplitudes, $\rho_{\parallel,\omega}$ is the sum of
the square moduli of the coefficients of $E_\parallel$ in eqs.
(\ref{erefl}). Similarly, $\rho_{\perp,\omega}$ is obtained by
adding together the square moduli of the coefficients of
$E_\perp$.

The absorption coefficient $\alpha_\omega=1-\rho_\omega$ has been computed
numerically in the relevant
angular ranges following the procedure outlined above and the results
have been used to evaluate the integral
in eq.~(\ref{df}). Although the numerical scheme is rather
straightforward,
care should be used since the refractive index becomes resonant
where the coefficient of the higher order term
in eq.~(\ref{disp}) vanishes (B80; see also \citealt{mel86}).
This happens at $P+v\sin^2\alpha=0$, that is to say at the two frequencies
\begin{equation}\label{resonan}
\omega^2_\pm = \frac{\omega_p^2+\omega_B^2}{2}\left\{1 \pm\left[1-
\frac{4\omega_p^2\omega_B^2\cos\alpha}{(\omega_p^2+\omega_B^2)^2}\right]^{1/2}
\right\}\, .
\end{equation}

Assuming no collisional damping has the main advantage that the
fourth order polynomial obtained by squaring the dispersion
relation has real coefficients and its roots are either real or
complex conjugates in pairs. Numerical experimenting shows that
the roots develop an imaginary part only close to the resonances
and are real and distinct otherwise. As discussed by
\citet{mel86}, there exist two cut-off energies at which one of
the two indices vanishes. The cut-off energies are
$\alpha$-dependent and the smallest one, $\omega_0$, falls in the
range $\omega_-<\omega_0<\omega_+$. This corresponds to the
appearance of an evanescent mode, which can not propagate into
the medium. Modes for which the refractive index has a large
imaginary part are severely damped (e.g. \citealt{ja75}), so they
can not penetrate much below the surface either. We point out that
the existence of these damped waves is not in contradiction with
having neglected collisional damping. The ``conductivity''
$\sigma^{pl}_{ij}$ of a cold plasma can be computed from the
dielectric tensor (\ref{maxw}) using the standard relation
\begin{equation}
\label{sigmaplasma}
\epsilon_{ij} = \delta_{ij} + i \frac{4 \pi }{\omega} \sigma^{pl}_{ij}
\end{equation}
and is therefore purely imaginary (see \citealt{ja75};
\citealt{mes92}). Quotation marks are placed on ``conductivity''
because there is no resistive loss of energy in this case. Yet,
depending on their frequency and grazing angle, electromagnetic
waves may be exponentially damped in the cold electron plasma.

In all cases in which one of the acceptable roots is real and
negative or it is complex and its imaginary part exceeds a given
value, only one refracted mode survives. Consequently, we adopt a
``one mode'' description to derive the reflectivity. By
specializing the previous calculation to a single mode (labeled
`1' for convenience), and defining $\cc_1 = \epsilon_{31} a_1 +
\epsilon_{32} b_1 + \epsilon_{33}$, we obtain (see also
\citealt{ja75})

\begin{equation}\label{onemode}
E''_\parallel =  \frac{\cc_1 \arctan i - \ac_1}{ \cc_1 \arctan i + \ac_1}
E_\parallel\,
,\quad
E''_\perp  = \frac{\sqrt{n^2_1 -\sin^2 i}- \cos i}{\sqrt{n^2_1 -\sin^2 i}
+ \cos i} E_\perp
\end{equation}
from which the reflectivity follows.

As we mentioned earlier, selecting the two relevant roots of the
fourth order polynomial may become difficult in some parameter
ranges. While for most values of the energy and angles, the two
meaningful roots produce a residual many orders of magnitude
below that of the spurious ones, in some cases all residuals are
small and of the same order. We encountered situations in which,
owing to round-off errors and despite the use of
quadruple-precision complex arithmetic, one of the relevant roots
produced a residual larger than that of the spurious solutions. A
bad choice of the roots usually gives negative absorption. In
these cases the calculation is repeated with a different choice
of the roots until $\alpha_\omega$ is positive and its value
close to those computed for neighboring values of the parameters.

The quantity $f_\omega/B_\omega(T)$ (see eq.~[\ref{df}]) is shown
in Figure~\ref{alpha_ang} as a function of energy for
$B=10^{12}$~G and different values of the angle $\alpha$. Here the
surface temperature is assumed constant. In general, the emissivity
becomes lower as the magnetic field is more and more tilted with
respect to the surface normal. The strong absorption around the
resonant frequency $\omega_-$ is clearly seen; the absorption dip
becomes more pronounced (and the surface emissivity decreases) as
$\alpha$ approaches $\pi/2$. The total monochromatic emissivity is
obtained integrating over the entire stellar surface
(eq.~[\ref{ftot}]), once the magnetic field topology is specified.
Here and in the following we assume that the field is dipolar,
$B= B_p[(4-f)\cos^2\theta+f]^{1/2}/2$, where $B_p$ is the polar
field strength and $f\simeq 1.2$ accounts for
general-relativistic corrections in a Schwarzschild space-time
(see e.g. \citealt{pz2000}). Accordingly, $\alpha$ is related to
the magnetic co-latitude $\theta$ by $\cos^2\alpha=
\{4-4f/[(4-f)\cos^2\theta+f]\}/(4-f)$.

The quantity $F_\omega/B_\omega(T)$ is shown in
Figure~\ref{alpha_tot} for different values of $B_p$ (models
computed accounting for $e$-phonon interactions are also shown,
see \S~\ref{damp} for details). Again, the surface temperature is
taken constant. As expected, integration over $\theta$ smears out
any strong feature around $\omega_-$, as it can be seen comparing
the full line in Figure~\ref{alpha_tot} with those in
Figure~\ref{alpha_ang} (for $B_p > 10^{12}$ G the angle-averaged
resonant feature lies outside the energy range that we have
considered). Below $\omega_-$ one of the two modes is
non-propagating, a so-called whistler. Whistlers have a very
large refractive index, which diverges at $\omega_-$ and for
$\omega\to 0$, and this explains the high reflectivity at these
frequencies (e.g. \citealt{mel86}). The increase of the plasma
frequency with $B$ is responsible for the lower emissivity at
larger fields and, if we restrict to energies below the
(angle-averaged) resonance $\omega_-$, the dependence on energy is
about the same (i.e. the curves are nearly self-similar, see
again Figure \ref{alpha_tot}).

The model presented so far has been computed using for the plasma
frequency in the surface layers our reference value
$\omega_{p,0}$. It should be stressed, however, that this value,
which relies on the zero-pressure surface density given by
eq~(\ref{dens}), is just an estimate (see e.g.
\citealt{lai2001}). In order to assess the effects of this on the
emitted spectrum, we explored the parameter space by varying the
electron plasma frequency around $\omega_{p,0}$; this accounts
also for the uncertainties on the number of free electrons per
nucleon, i.e. deviations from our reference value $Z=26$.

While the energy dependence of the absorption coefficient does not change
significantly, an increase in the plasma
frequency at fixed $B$ produces an overall decrease of the emissivity.
This is illustrated in Figure~\ref{reduc},
where the ratio of the emitted to the blackbody power in the 0.1-2 keV band
is plotted against the plasma frequency for $B_p=3\, ,5 \times 10^{13}$~G.
As can be seen from
Figure~\ref{reduc}, in this simple description we expect that the star
surface radiates only $\sim 30\%$ of the blackbody
power if the plasma frequency (the density) is about a factor 6 (36)
higher than the estimate provided by eq. (\ref{omegap}) (eq.
[\ref{dens}]). Also, since in this simplified
description $f_\omega/B_\omega(T)$ is found to vary with the
magnetic field angle (Figure~\ref{alpha_ang}),
viewing effects may be relevant with even larger depression expected if
the star is viewed equator-on.

\subsection{Electron-phonon Interaction}
\label{damp}

In the calculation presented so far all collisional damping
effects have been neglected, therefore the only source of
``conductivity'' is the cold plasma.
This is of course an oversimplification. In the solid crust of a
NS with $B\sim 10^{12}-10^{13}$~G and $T \sim 10^6$~K electrons
are strongly degenerate ($T \ll T_F$, where $T_F$ is the Fermi
temperature). Furthermore, if $\rho \sim \rho_s$ quantum effects
due to the magnetic field are not negligible. In the degenerate
surface layers charge (and heat) is transported primarily by
electrons. Several efforts have been devoted to an accurate
determination of the electrical conductivity $\sigma$, mostly
because this is the basic quantity governing the magnetic field
evolution (see e.g. \citealt{flit76}; \citealt{yakur80};
\citealt{pot99} and references therein). In particular, it is
well known that while at temperatures above the crystallization
temperature of the ions the main factor governing the electrical
conductivity is scattering off ions, below the crystal melting
point the dominant process is scattering by crystal lattice
vibrations (phonons) through Umklapp processes. Eventually, if
the temperature decreases further, impurities in the crystal
structure and scattering off lattice defects start to be
important.

Magnetic fields in the neutron star crust complicate electron
transport, in particular, making it anisotropic (\citealt{kaya81};
\citealt{yak84}; \citealt{hern84}; \citealt{pot99}). Electrons
move freely parallel to $\mathbf{B}$, but their motion
perpendicular to the field is quantized. Therefore, transport
properties will be affected through the influence of the discrete
spectrum on the density of states and collision times. The
electrical conductivity tensor is computed from the transport
tensor and can be expressed as (\citealt{hern84}; \citealt{pot99})
\begin{equation}
\label{sigint}
\sigma_{ij} = \int e^2 \frac{ {\it N}_B \left(\varepsilon \right )
}{\varepsilon/c^2} \tau_{ij} \left(\varepsilon \right ) \left ( -\frac{
\partial f_0 }{\partial \varepsilon } \right ) d \varepsilon \, ,
\end{equation}
where $\varepsilon$ is the electron energy (including the rest energy
$m_e c^2$), $f_0
\left(\varepsilon \right )$ is the Fermi-Dirac distribution,
\begin{equation}
 {\it N}_B \left(\varepsilon \right ) = \frac{m_e \omega_B }{2 \left ( \pi
\hbar \right )^2} \sum_{n=0}^{n_{max}} g_n \left [ \left (
\varepsilon/c \right )^2 -  \left (m_e c \right )^2 - 2 m_e
\hbar \omega_B n \right ]^{1/2} \, ,
\end{equation}
$g_n$ is the statistical weight of the $n^{th}$-Landau level and
$n_{max}$ is the maximum Landau number for a given energy
$\varepsilon$ (see \citealt{pot99} for all details). The
quantities $\tau_{ij} \left(\varepsilon \right )$ in
eq.~(\ref{sigint}) play the role of relaxation times. The tensor
$\sigma_{ij}$ has three independent components: one parallel to
the field ($\sigma_{zz}\equiv\sigma_\|$), one transverse
($\sigma_{xx}=\sigma_{yy} \equiv \sigma_\perp$) and one
off-diagonal (Hall) component $\sigma_{xy}\equiv\sigma_H$, which
is non-dissipative. In general evaluating eq.~(\ref{sigint})
requires energy-integration, but for strongly degenerate
electrons and not too close to the Landau thresholds this becomes
unnecessary. In this case it is $\sigma_{ij} \approx \left (e^2
n_e c^2 /\varepsilon_F \right) \tau_{ij}\left(\varepsilon_F
\right )$ where $\varepsilon_F$ is the Fermi energy.

The way in which the magnetic field affects the charge transport
depends on its strength. According to \cite{pot03}, the
non-quantizing (classical) regime occurs if $T\gg T_B$, where
$kT_B = \hbar e B / (m_e c)$. Even a non-quantizing magnetic
field, which essentially does not affect the thermodynamic
properties of matter, hampers transverse motion and produces Hall
currents. In the opposite case, $T\ll T_B$, one can distinguish
between a weakly quantizing (when electrons populate several
Landau levels) and a strongly quantizing regime (when the
magnetic field confines most electrons in the ground Landau
level). Transition to the strongly quantizing regime occurs below
the first Landau threshold, i.e. for $\rho < \rho_B \approx 7
\times 10^3 (A/Z) B_{12}^{3/2}$~g/cm$^3$. Note that the zero
pressure density $\rho_s$ is much smaller than $\rho_B$ for $B
\gg 10^{10}$~G.

So far, the most complete expressions for the electrical
conductivity are those computed by \cite{pot99}, by taking into
account correlation effects in the strongly coupled Coulomb liquid
and multi-phonon scattering in the Coulomb crystal
(\citealt{baiko98}). These results show that in a weakly
quantizing field the conductivity in the longitudinal and
transverse direction oscillates around their classical values.
However, such oscillations are quite prominent in the regime of
strong quantization, where they may reach several orders of
magnitudes. In this case the transport properties of the matter
are very different from those in the classical regime.

In order to include effects of electron-phonon scattering in our
computation, it is convenient to introduce the effective relaxation times.  
In the regime of strongly degenerate electrons, when thermal averaging is
unimportant, the relaxation times are related to the conductivity along
the various directions by $\tau_{xx}=\tau_{yy}=
4\pi\sigma_\perp/\omega_p^2$, $\tau_{zz}=4\pi\sigma_\|/\omega_p^2$ (e.g.
\citealt{ziman}; \citealt{yakur80}; \citealt{pot99}). Then we proceed in a
standard (although approximate) way by summing the relevant effective
frequencies of the different processes. Note that, while for transport 
along the field the effective collision frequency for electron-phonon 
damping is simply $\sim\tau_\|^{-1}=\tau_{zz}^{-1}$, for transport
across the field it is not always $\sim\tau_{xx}^{-1}$. In the strong 
field regime it becomes directly proportional to $\tau_{xx}$. 
Physically this reflects the fact that the relaxation time in the
transverse direction is, in the strong field regime, longer than 
the time between electron-phonon collisions ($\tau_\perp$) which 
in turn determines the damping according to Heisenberg's principle. 
Accordingly, we derived the collision frequency 
from the interpolation formula by \cite{pot99}, valid at any field
strength $\tau_{xx} = \tau_\perp/(1+\omega_B^2\tau_\perp^{-2})$.

We focus on the case $\mathbf{B} \parallel z$
(i.e. $\alpha=0$) and write the conductivity tensor of the pure
plasma component (eq.~[\ref{sigmaplasma}]) in rotating coordinates
$\left (\mathbf{e}_+ , \mathbf{e}_-, \mathbf{e}_z \right)$,
$\mathbf{e}_\pm = \mathbf{e}_x \pm i \mathbf{e}_y $
(\citealt{mes92}). In this frame the conductivity tensor $\tilde
\sigma^{pl}$ is diagonal\footnote{ The plasma polarization
tensor, whose elements are directly related to the collision
times, is in fact $\Pi_{ij} \equiv v^{-1} \left (\delta_{ij} -
\epsilon_{ij} \right ) \equiv - (i 4 \pi / v \omega) \sigma_{ij}$,
where $v=\omega_p^2/\omega^2$. In a magnetic field and in
rotating coordinates with $\mathbf{B}\parallel z$, the polarization
tensor is diagonal (see e.g. \citealt{mes92}).}
\begin{equation}
\tilde \sigma^{pl}_{ij} = \Lambda_{ik} \sigma^{pl}_{kl}
\Lambda^{-1}_{lj} = \frac{i \omega_p^2}{4 \pi \omega}
\left(\begin{array}{ccc}\displaystyle \frac {1}{1 +
\omega_B/\omega} & 0 & 0 \cr 0 & \displaystyle\frac {1}{1 -
\omega_B/\omega} & 0 \cr 0 & 0 & 1
\end{array}\right) \,,
\end{equation}
where
\begin{equation}
\Lambda_{ij} = \left(\begin{array}{ccc} 1 & i & 0 \cr 1 &  -i  & 0
\cr 0 & 0 & 1
\end{array}\right) \,.
\end{equation}
The conductivity tensor related to electron-phonon scattering can be
written (again in rotating coordinates) as
\begin{equation}
\tilde \sigma^{e-ph}_{ij} = \left(\begin{array}{ccc} \sigma_\perp
& 0 & 0 \cr 0 &  \sigma_\perp & 0 \cr 0 & 0 &
\sigma_{\|}
\end{array}\right) \,.
\end{equation}

We then obtain 
\begin{eqnarray}
\tilde \sigma_{zz}^{tot}& =& \left(\frac{\omega_p^2}{4\pi}\right)
\left [\left (\frac{4\pi\tilde \sigma_{zz}^{pl}}
{\omega_p^2}\right )^{-1} + \tau_{\|}^{-1} \right ] ^{-1} = \frac{i
\omega^2_p}{4 \pi \omega  } \frac{1}{1 + i \omega^D_{\|}/\omega}\, ,\cr
\tilde \sigma_{xx}^{tot} & = & \left(\frac{\omega_p^2}{4\pi}\right)
\left [ \left (\frac{4\pi\tilde 
\sigma_{xx}^{pl}}{\omega_p^2}\right )^{-1} + \tau_{\perp}^{-1} 
\right]^{-1} =
\frac{i  \omega^2_p}{4 \pi \omega  } \frac{1}{1+ \omega_B
/\omega + i \omega^D_{\perp}/\omega} \, ,  \cr
\tilde \sigma_{yy}^{tot} & = & \left(\frac{\omega_p^2}{4\pi}\right)
\left [ \left (\frac{4\pi\tilde \sigma_{yy}^{pl}}{\omega_p^2}
\right)^{-1} + \tau_{\perp}^{-1} \right ] ^{-1} = \frac{i
\omega^2_p}{4 \pi \omega  } \frac{1}{1 -  \omega_B /\omega + i
\omega^D_{\perp}/\omega} \, ,
\end{eqnarray}
where the quantities
\begin{eqnarray} 
\omega^D_{\|}& =& \frac{\omega_p^2}{4\pi\sigma_{\|}}\, , \cr
\omega^D_{\perp}& =& \frac{1}{2}\left(\frac{\omega_p^2}{4\pi\sigma_{\perp}}
\right)\left[1-\sqrt{1-4\omega_B^2\left(\frac{4 \pi\sigma_\perp}
{\omega_p^2}\right)^2}\right]
\end{eqnarray}
play the role of damping frequencies in the two
different directions. By transforming back to non-rotating
coordinates we finally get
\begin{equation}
\frac{i  4 \pi }{\omega  } \sigma^{tot}_{ij} = \frac{1}{2}
\left(\begin{array}{ccc} L^{tot} + R^{tot} - 2 &
 i \left(  L^{tot} -  R^{tot} \right)   & 0 \cr
-i \left(  L^{tot} -  R^{tot} \right)   &   L^{tot} +
R^{tot}
- 2
& 0 \cr
0 & 0 & 2 P^{tot} -1
\end{array}\right) \,,
\end{equation}
from which the dielectric tensor follows as
\begin{equation}
\epsilon^{tot}_{ij} \equiv \delta_{ij} + \frac{i  4 \pi }{\omega
} \sigma^{tot}_{ij} = \left(\begin{array}{ccc} S^{tot} &
 - i D ^{tot}   & 0 \cr
i D^{tot}  &   S^{tot}
& 0 \cr
0 & 0 & P^{tot}
\end{array}\right) \,.
\end{equation}
In the previous expressions, the off-diagonal terms correspond to the 
(non-dissipative) Hall conductivity, and $S^{tot}$, $D^{tot}$ are defined 
as $S,D$ in \S~\ref{cold} but with $R,L,P$ replaced by
\begin{equation}
\label{rtlt}
\left(\begin{array}{c}R^{tot} \cr L^{tot} \end{array}\right)
= 1-\frac{\omega_p^2}{\omega^2}\frac{\omega}{\omega\mp\omega_B
+ i \omega^D_{\perp}}
\, ,
\end{equation}
\begin{equation}
\label{pt} P^{tot} =1-\frac{\omega_p^2}{\omega^2}
\frac{\omega}{\omega + i \omega^D_{\|}} \, .
\end{equation}
A further rotation by an angle $\alpha$ accounts for the
misalignement between $\mathbf{B}$ and $z$ and gives the
dielectric tensor $\epsilon^{tot}_{ij}$  in the same form as in
eq.~(\ref{maxw}).

We have repeated the computation of the monochromatic absorption
coefficients by following the same method as in \S~\ref{cold},
but using the dielectric tensor $\epsilon^{tot}_{ij}$. The conductivites 
$\sigma_{\|, \perp}$ have been
computed numerically for the appropriate values of $B$, $\rho$
and $T$ \footnote{ The package CONDUCT.FOR developed by A.
Potekhin and available at
www.ioffe.ru/astro/conduct/condmag.html has been used.}. 
The inclusion of electron-phonon damping seriously affects the 
emission properties of the surface, as can be seen from 
Figures~\ref{alpha_tot} and \ref{alpha_ang_phon} where the angle-averaged 
and angle-dependent emissivity is shown. 

Below $\sim 1$~keV the emissivity declines quite rapidly with 
decreasing photon energy. There is
now a strong dependence on the magnetic field strength, with the 
suppression 
being more pronounced at lower fields. This is mainly due to the
role played by electron-phonon damping in the transverse plane which
is quantified by the real parts of $\sigma_{xx}^{tot}$, $\sigma_{yy}^{tot}$.
These are $\approx \omega_p^2 \omega_\perp^D/(4 \pi \omega_B^2)$ and increase
monotonically with decreasing $B$. In particular, while for $B_p\sim 5 
\times 10^{13}$~G the star surface can radiate down to a few tens of eV's,
for $B_p\lesssim 5 \times 10^{12}$~G, a sharp edge appears at $\sim
300$~eV and the star surface behaves as a perfect reflector at energies
below $\approx 100$~eV. An absorption feature close to the cutoff
energy $\omega_0$ is now
more clearly seen at $\approx 300$~eV for the field strengths reported
here. Contrary to the behaviour of the resonant frequency $\omega_-$, 
the value of the cutoff energy
changes only very weakly with $B$ (see Fig.~\ref{alpha_ang_phon}). This
implies that the feature survives even after integration over the entire  
surface has been performed (see Fig.~\ref{alpha_tot}).
As in the cold electron gas
considered in \S~\ref{cold}, the emissivity is strongly
dependent of the angle between the magnetic field and the surface
normal (see again Fig.~\ref{alpha_ang_phon}), so
line-of-sight effects are expected to be important.

The different behaviour produced by the inclusion of damping can
be understood as follows. We start rewriting the dispersion
relations (eq.~[\ref{disp}]) in terms of the
angle of refraction $\Theta$, and then compare the refractive
indices computed with and without the damping terms.
We consider the case $ \mathbf{B}\parallel z$ (i.e.
$\alpha=0$; see \citealt{mel86}). The ensuing equation is quadratic
in $n^2$ and therefore simpler to solve. Results are shown in
Figure~\ref{n1n2} for $\Theta \sim 43^\circ$ and two different magnetic
field strengths. As can be seen from the bottom panel
of Figure~\ref{n1n2},
when collisional damping is included, the $n_m^2$ develop a substantial
imaginary part over a wide range of energies which can easily
extend up to an order of magnitude below $\omega_-$. The region of
interest is larger for decreasing field strength and/or increasing 
grazing angle. This causes the
angle-dependent absorption feature that is seen in
Figure~\ref{alpha_ang_phon} at $\log E \sim -0.6, 0$ and large $\alpha$.
Moreover, below a certain energy threshold
both indices develop a large imaginary part so that both
propagation modes are substantially damped. Since the penetration
depth is $\delta = c/[\omega {\rm Im}(n)$], in our
angle-dependent computation we neglect the contribution of modes
with ${\rm Im}(n) = \lambda /(2 \pi \delta) > 0.01$ (dashed line
in Figure~\ref{n1n2}).

Of course the most important consequence is that the results now depend
in a crucial way on the choice of the rejection limit, which in turn is
only one of the aspects of the uncertainties in the physics
at the vacuum/surface interface. The relevance of varying the adopted limit
for rejecting on the surface emissivity is shown in
Figure~\ref{rejec}.


\section{Discussion}
\label{discuss}

This investigation has been motivated by recent X-ray
observations of \oneight \ and \zeroseven, the two brightest
among the seven isolated neutron stars discovered by {\it
ROSAT\/}. In particular, detailed {\it Chandra\/} observations
have convincingly shown that \oneight \ has a featureless thermal
spectrum for which a simple blackbody distribution seems to
provide a better fit to X-ray data than more sophisticated
atmospheric models (\citealt{bur2001}; \citealt{dra2002}).
\oneight \ has a firmly established optical counterpart
(\citealt{wm97}; \citealt{vkk2001a}, similarly to \zeroseven \
(\citealt{mh98}; \citealt{kvk98}). Accurate spectroscopy and
photometry with combined Very Large Telescope (VLT) and Hubble
Space Telescope (HST) data has shown that the UV-optical energy
distribution closely follows a Rayleigh-Jeans tail
\citep{vkk2001a}. However, as originally noticed by \cite{wm97},
the Rayleigh-Jeans tail of the X-ray best-fitting blackbody
underpredicts the optical flux by about a factor 6
\citep{wl2002}, and this has been taken as suggestive of emission
from regions on the star surface with different properties (see
below). Deep HST observations have also revealed the presence of
a bow-shock nebula in H$_\alpha$ around \oneight \
\citep{vkk2001b}. \cite{w2001}, by means of HST observations,
derived the proper motion and parallax of the star, and obtained
a distance of about 60 pc.  However, this distance was shown to
be in error by recent re-analysis of the same HST data
\citep{kka2001}, and of these data augmented by further
observations \citep{wl2002}.  These studies place the source at
about 120~pc, or twice the original distance.  The simultaneous
determination of the distance and X-ray flux (under the
reasonable assumption that it comes from the stellar surface)
makes \oneight \ unique in its class inasmuch it allows a direct
estimate of the radiation radius
\begin{equation}\label{radradius}
R_\infty = 4.25\, \left(\frac{D}{100 \, {\rm pc}}\right)
\left(\frac{T_{bb}}{60\, {\rm eV}}\right)^{-2}\, {\rm km}\, .
\end{equation}

Based on the location of \oneight \ in front of the R~CrA molecular
cloud, \cite{dra2002} have shown that their derived neutral H column
density of $1\times 10^{20}$~cm$^{-2}$ limits the distance to $\leq
170$~pc.  Taken at face value,
the expression for the radiation radius
above yields for this distance a value of at most $\sim 8.2$~km.
Such a figure is
incompatible with current bounds on the stellar radius (as
measured by an observer at radial infinity) based on theoretical
investigations of the equation of state of matter at ultra-high
densities (EOSs; see, e.g., \citealt{lapra2001}), $12\, {\rm km}\lesssim
R_\infty \lesssim 17\, {\rm km}$. This discrepancy motivated the
suggestion that \oneight \ might be a strange/quark star
(\citealt{hae2001}; \citealt{xu2002}; \citealt{dra2002};
\citealt{gon2002}).

More conventional scenarios involving a neutron star have been discussed in
connection with \oneight.
\cite{pons2002} explored non-magnetized model atmospheres with
different compositions (H, He, Fe, Si-ash) in order to reproduce
the emission properties of \oneight.
H/He spectra are almost featureless but they deviate substantially from a
blackbody, showing an excess at higher energies. As was
already noted by \cite{cam97} and further remarked on by
\cite{pons2002}, the fit with H/He atmospheric models yields a
column density distance $\lesssim 10$ pc for standard values of
the star radius, more than one order of magnitude below the
parallax measurement.

\cite{pons2002} considered both a uniform thermal distribution and a
two-temperature surface model.  For the adopted distance of 61 pc
\citep{w2001}, they concluded that the apparent radius is $\approx
7$--8 km for a uniform thermal distribution---too small for any EOSs.
Owing to the reduced area of the hotter X-ray emitting region,
two-component models provide larger values for $R_\infty$ and may
explain also the excess at optical wavelengths over the best-fitting
X-ray blackbody. With a revised distance of 117 pc, \cite{wl2002}
argued that both the single-component heavy-element (Fe and Si) and
two-component blackbody models of \cite{pons2002} yield acceptable
values for the stellar radius.  In particular, they claim that a
two-temperature model in which a blackbody at $T=15$ eV is emitted by
a region with an angular diameter five times larger than the X-ray
blackbody ($T=63$ eV) can reproduce the multiwavelength SED.

These conclusions are not without problems.  Although detailed
spectral calculations were announced, the present results of
\cite{wl2002} rely on the assumption that the two regions on the
stellar surface emit a pure blackbody spectrum. How to justify
this assumption for a neutron star, however, remains
unexplained.  Moreover, the presence of a small, hot region on
the star surface might be difficult to reconcile with the lack of
pulsations in the X-ray flux, as stressed by \cite{dra2002},
especially in the light of the present very tight limits on the
pulsed fraction of this source ($\lesssim 1.3\, \%$; \citealt{bur2003}).
All heavy element
spectra calculated by \cite{pons2002} exhibit a variety of
emission/absorption features in the soft X-ray range.  No
evidence for such features is present in the {\it Chandra\/} 
and {\it XMM-Newton\/} data.

More recently
\cite{braro2002} readdressed the two-temperature surface model
for \oneight, pointing out that several effects (isothermality,
magnetic smearing, rotation) may act in suppressing the spectral
features from an extended atmosphere with heavy elements. In
particular, they discuss in detail the role of rotation, showing
that phase-dependent Doppler shifts in a rapidly rotating neutron
star ($P\approx 1$ ms) wash out all features, leaving a nearly
planckian spectrum. Although such a short period can not be
excluded on the basis of present data, the detected periods of
other thermally emitting INSs are in the range $\approx$ 0.1--10
s, about two orders of magnitude larger. In the \cite{braro2002}
picture the X-ray emitting region is kept warm by external
heating, and the genuine surface temperature should correspond to
the cooler blackbody at $T\sim 15$ eV. A millisecond period appears hardly
compatible with the star's age of, as implied by conventional cooling
curves, $\approx 10^6$ yr. Furthermore, the energetics of the
bow-shock nebula implies $P = 4.6 \left ( B/10^8~{\rm G}
\right)^{1/2}$~ms.  A ms spin period therefore necessarily
demands for a very low field star. Such a low field seems hard to
reconcile with the limit on the age derived again from the
bow shock energetics, $(B/10^{12}~{\rm G})(\tau/10^6~{\rm yr})
\sim 3-4$.

Although two-temperature models appear promising in explaining the
multiwavelength SED of \oneight, no conclusive evidence has been
provided yet that a near blackbody, featureless spectrum can be
emitted by an extended atmosphere covering the stellar crust. An
alternative possibility, originally suggested by \cite{bur2001}
(see also \citealt{bur2003}) and further explored here, is that
\oneight \ may be a solid-surface NS. If this is the case, a
severe reduction in the surface emissivity has to be expected at
energies below the plasma frequency, according to the analysis of
\cite{trule78} and \cite{bri80}. The bare NS model may ease the
radiation radius problem. In fact, denoting with $f_E$ the ratio
of emitted to blackbody power and assuming emission from the
entire star surface, the value of $R_\infty$ now contains an
additional $f_E^{-1/2}$ factor with respect to that given by
eq.~(\ref{radradius}). As expected, the reduced surface
emissivity acts precisely in the same way as a reduced emitting
area, requiring a larger star radius. In order to represent a
viable option, the bare NS picture must conform to three basic
requirements: i) the conditions for the appearance of a solid
phase should be met, at least within the present uncertainties;
ii) the X-ray spectrum emitted by the surface should be very
close to blackbody in the 0.1--2 keV range to match {\it
Chandra\/}/{\it XMM} observations; and iii) quite low values of
$f_E$ ($\approx 0.1$) should be possible for $R_\infty$ to be in
the range allowed by current EOSs.

\begin{figure}
\includegraphics[width=6in,angle=0]{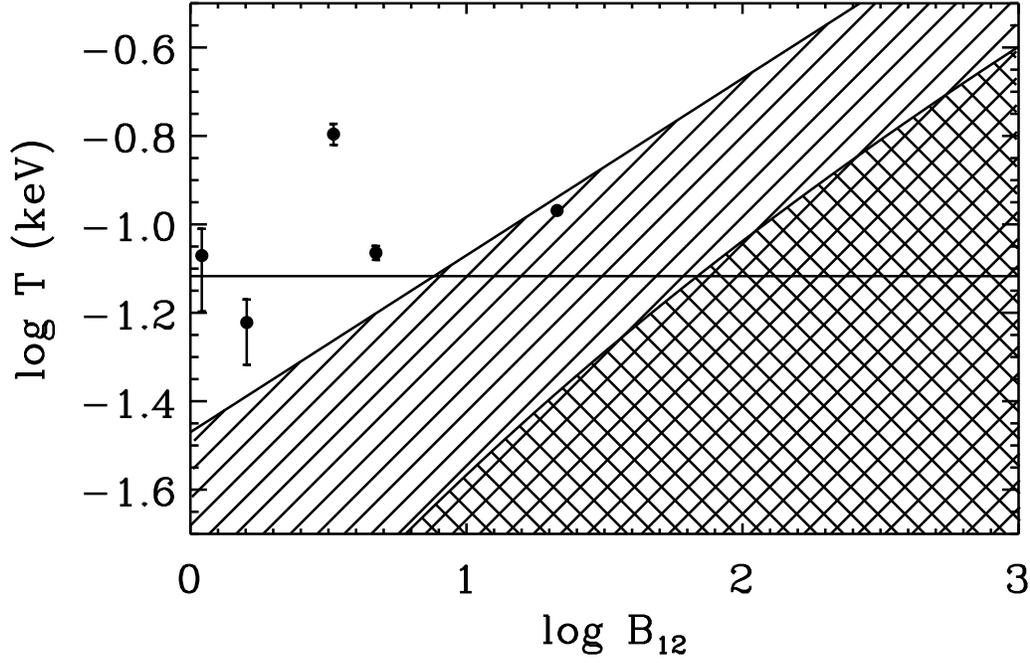}
\caption{\label{tcrit}The critical temperature for H and
Fe as a function of the magnetic field. Condensation is
possible in the shaded region for Fe and in the cross-hatched region for
H. The full circles with error bars mark the position
of five cool, isolated neutron stars (see table \ref{tableins}). The horizontal
line is drawn in correspondence to the color temperature of \oneight.
}
\end{figure}

\begin{figure}
\includegraphics[width=6in,angle=0]{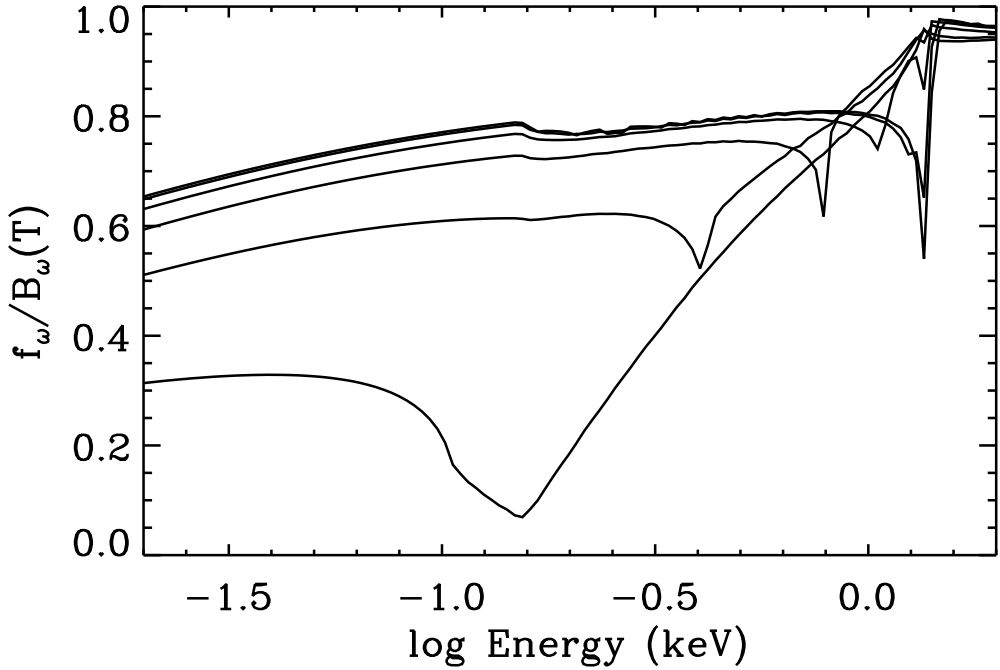}
\caption{\label{alpha_ang}The monochromatic absorption coefficient
as a function of energy for $B=10^{12}$ G and different values of the
magnetic field angle: curves are from top to bottom for
$2\alpha/\pi=0.05\, , 0.2\, ,0.4\, ,0.6\, , 0.8\,, 0.95$.
The plasma
frequency given by $\omega_{p,0}$ has been used here.
}
\end{figure}

In the absence of a measured period and period derivative, the
magnetic field of \oneight \ is still a mystery. Given a surface
temperature of $\sim 70$~eV, the magnetic field of \oneight \
should be in excess of $10^{13}$ G, more probably at least
3-5$\times 10^{13}$ G, for its surface layers of to be in the
form of condensed iron (see \S~\ref{bare} and Figure~\ref{tcrit}).
Although rather high, such a field strength is well below the
magnetar range and is noticeably shared by another {\it ROSAT\/}
isolated neutron star: \zeroseven \ \citep{silvia2002}. So,
although no definite conclusion can be drawn, the possibility
that \oneight \ is a solid-surface NS is real.

In the light of the results presented in \S~\ref{emission}, the
remaining two points are much more of an issue. When computed
accounting only for the  cold electron gas (\S~\ref{cold}),
spectra show indeed only small departures from a blackbody in the
0.1--2~keV band. A typical example is shown in
Figure~\ref{spec3e13}, where the computed spectra are plotted
together with the best-fitting blackbody (again in the 0.1--2~keV
range) for two different temperature distributions on the star
surface. Deviations from a blackbody are below 15--20\%, and this
would make possible to reasonably fit {\it Chandra} data. The
value of $f_E$ depends on the magnetic field and on the surface
density (see Figure~\ref{reduc}). In order to reach $f_E\sim 0.3$
(which produces an increase in radius of $\sim 2$) one must
invoke a density $\approx 50$ times larger than the zero pressure
value given by eq.~(\ref{dens}). Although the latter is only an
approximation, such a large departure might not be realistic. For
$\rho\sim \rho_s$, the predicted increase in radius is $\sim
15\%$, which is insufficient to reconcile the radiation radius
with canonical theoretical predictions, at least when viewing
angle effects are neglected (see also \citealt{thoma2003}).

\begin{figure}
\includegraphics[width=6in,angle=0]{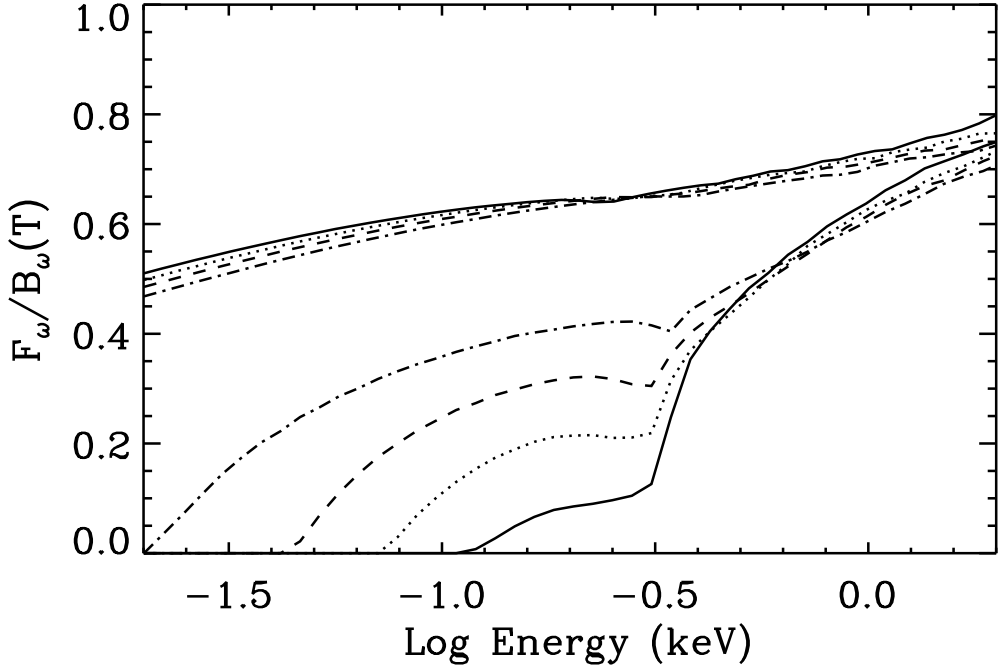}
\caption{\label{alpha_tot} The total absorption coefficient
averaged over the star surface, $F_\omega/B_\omega(T)$, as a
function of energy and for different values of the magnetic field:
$B_{p}=5\times 10^{12}$ G (full line), $ 10^{13}$ G (dotted
line), $2\times 10^{13}$ G (dashed line), $5\times 10^{13}$ G (dash-dotted
line). The two sets of curves correspond to models without and
with electron-phonon damping accounted for (the latter
are evaluated assuming $T=10^6$~K).
The plasma frequency $\omega_{p,0}$ has been used here. }
\end{figure}

\begin{figure}
\includegraphics[width=6in,angle=0]{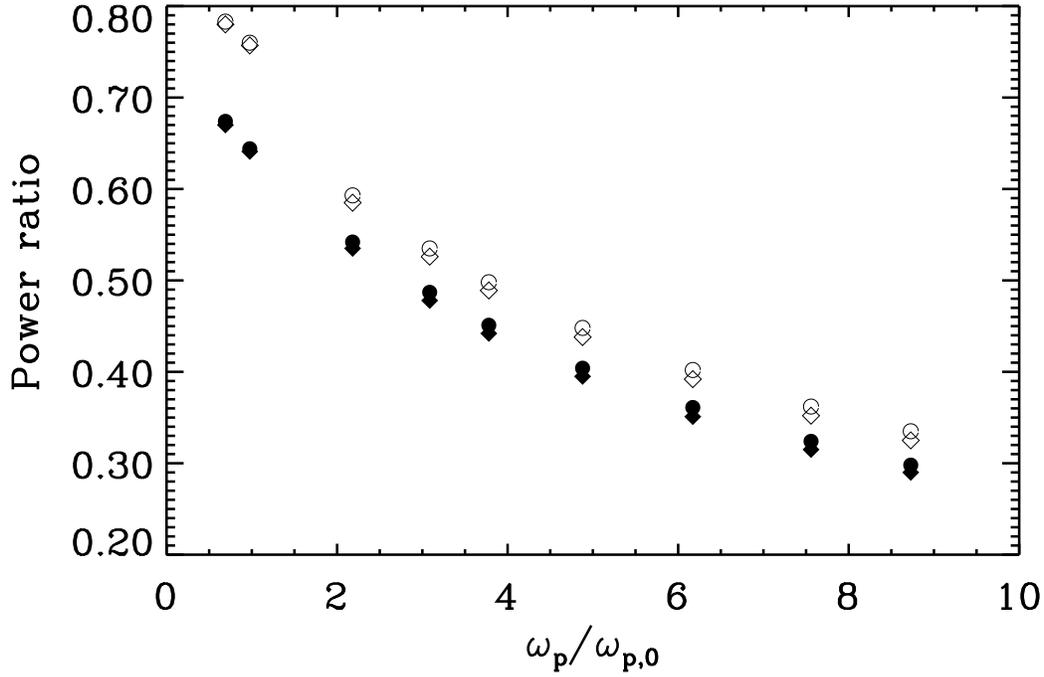}
\caption{\label{reduc}Ratio of the emitted to the
blackbody power in the 0.1--2 keV band
for different values of the plasma frequency. Circles refer to
$B_p=3\times 10^{13}$ G and diamonds to $B_p=5\times 10^{13}$~G.
Filled and open symbols are for the uniform/meridional variation
temperature distributions respectively. In the latter case we
assumed a profile $T^4(\theta) =
T^4_{surf}[K+(4-K)\cos^2\theta]/\{[1-0.47(1-K)]
(1+3\cos^2\theta)\}$ with $K=10^{-4}$ (e.g. \citealt{gh83};
\citealt{pmc96}).}
\end{figure}

\begin{figure}
\includegraphics[width=6in,angle=0]{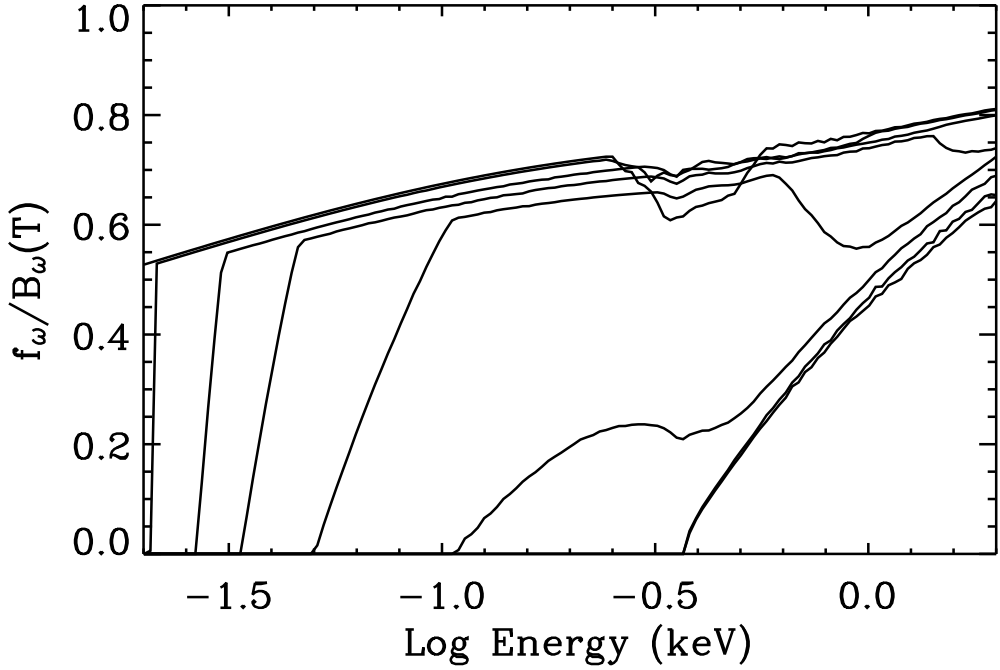}
\caption{\label{alpha_ang_phon}Same as in Figure~\ref{alpha_ang}, but with 
electron-phonon damping accounted for and $B=5 \times 10^{13}$~G, $T=10^6$~K;
curves are, from top to bottom for $2\alpha/\pi=0.05\, , 0.2\, ,0.4\, ,0.5\, ,
0.6\, , 0.7\, ,0.8\,, 0.95$.}
\end{figure}

\begin{figure}
\includegraphics[width=6in,angle=0]{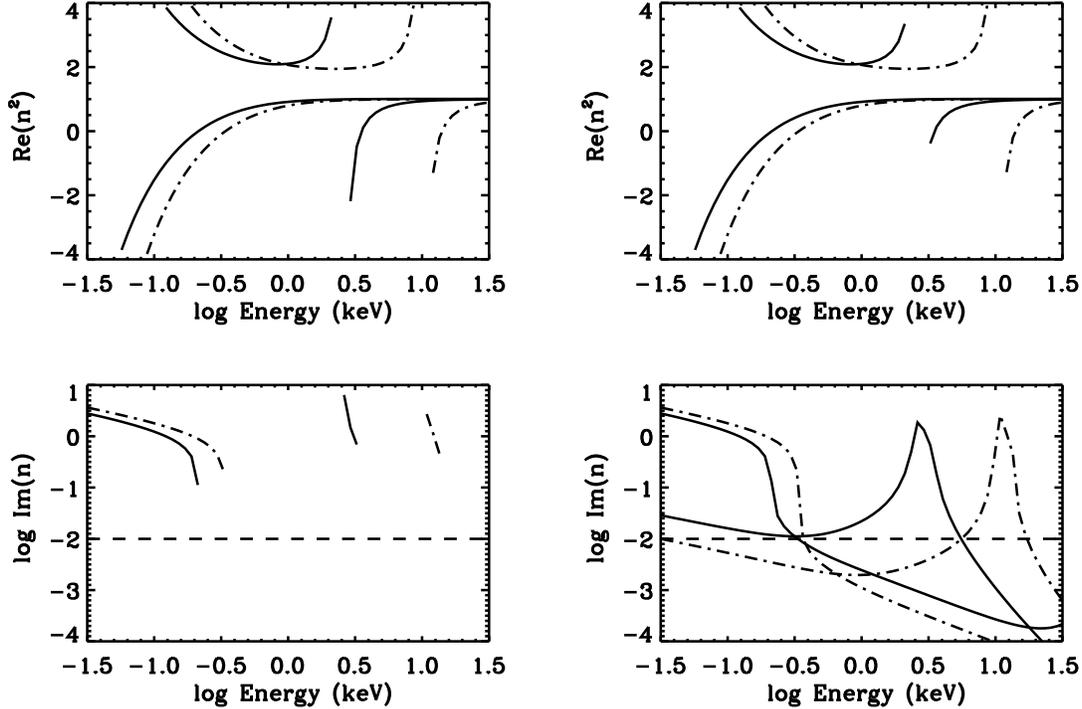}
\caption{\label{n1n2} Upper panels: ${\rm Re}(n_m^2)$ with (right) and without
(left) electron-phonon damping accounted for. The refractive indices are
the solutions of the dispersion relation for $\Theta \approx 43^\circ$,
$\alpha =0$, and $B=5 \times 10^{12}$~G (solid line), 
$B=5 \times 10^{13}$~G (dash-dotted line). Lower panels: same for the
imaginary parts of $n_m$. The dashed line represents our rejection criterion,
${\rm Im}(n)>0.01$ (see text).}
\end{figure}

\begin{figure}
\includegraphics[width=6in,angle=0]{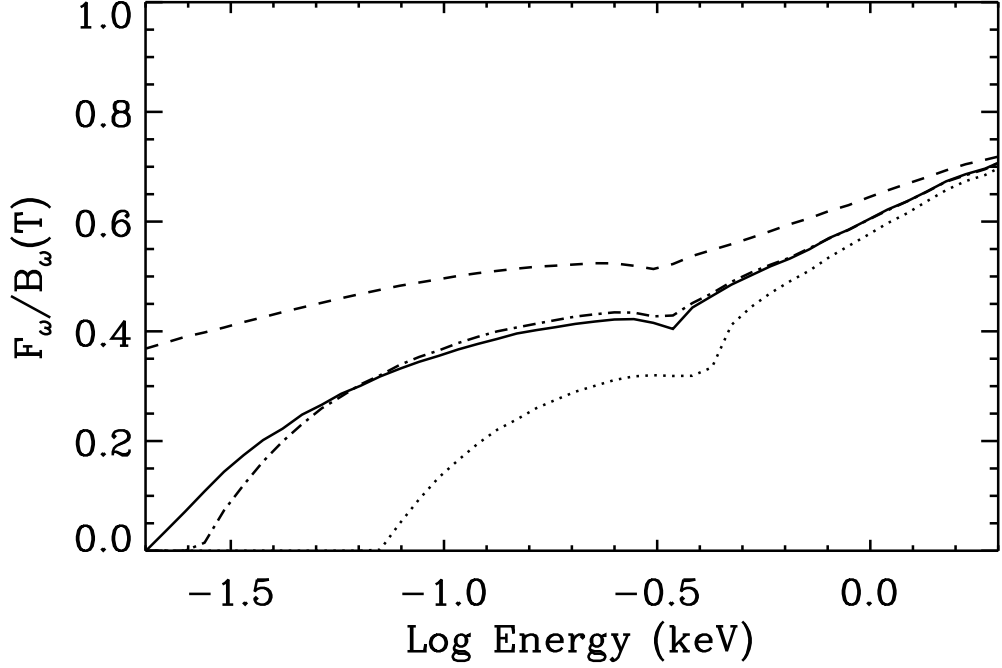}
\caption{\label{rejec} Same as in Figure~\ref{alpha_tot} for $B_p=5\times 
10^{13}$~G and different values of the rejection limit: ${\rm Im}(n)=0.01$
(solid line), ${\rm Im}(n)=0.05$ (dashed line) and ${\rm Im}(n)=0.005$
(dotted line). The dash-dotted line shows the surface emissivity
for the meridional temperature distribution (${\rm Im}(n)=0.01$; 
see the caption of Figure~\ref{reduc}) and is to be compared with the 
solid curve.}
\end{figure}

\begin{figure}
\includegraphics[width=6in,angle=0]{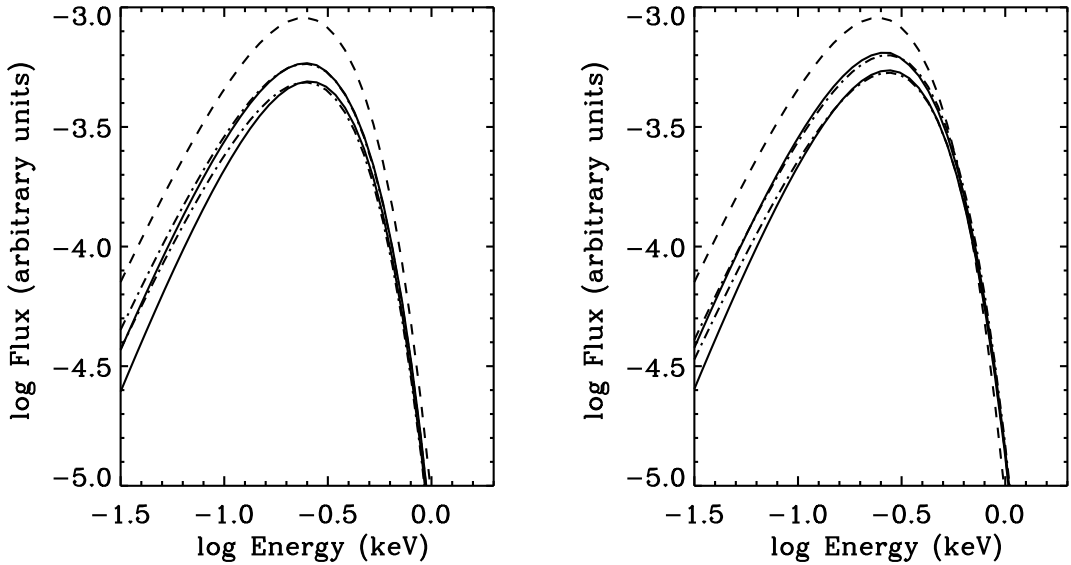}
\caption{\label{spec3e13}The emitted spectrum in the cold plasma
limit for $B_p=2\times 10^{13}$ G and $T_{surf}=10^6$~K. Left
panel: uniform surface temperature; right panel: meridional
temperature variation as defined in the caption of
Figure~\ref{reduc}. The dashed line is the blackbody at
$T_{surf}$ and the dash-dotted line the blackbody which best-fits
the calculated spectrum in the 0.1--2 keV range. The two models
shown in each panel are computed for $\omega_p=\omega_{p,0}$
(upper solid curve) and $2.45\omega_{p,0}$ (lower solid curve).
Spectra are at the star surface and no red-shift correction has
been applied.}
\end{figure}

\begin{figure}
\includegraphics[width=6in,angle=0]{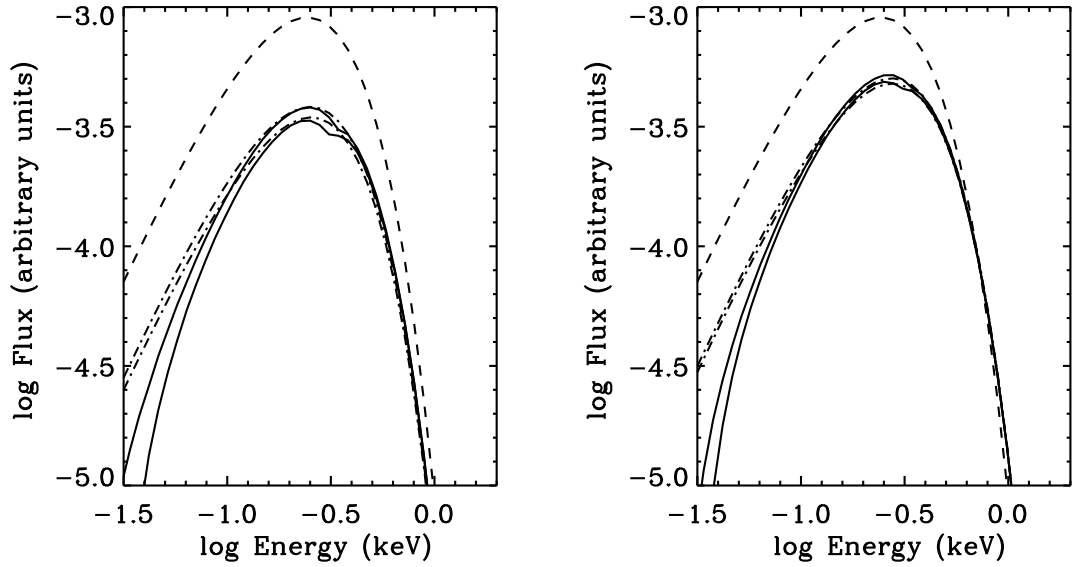}
\caption{\label{specdamp}Same as in Figure~\ref{spec3e13} for the case
with electron-phonon damping. The two models shown in each panel are 
computed for $B_p=2\times 10^{13}$ and $5\times 10^{13}$~G; here
$\omega_p=\omega_{p,0}$ and other details are as in Figure
\ref{spec3e13}.}
\end{figure}

These models are, however, unrealistic. The cold electron gas
assumption is a poor approximation at low energies where damping
by electron-lattice interactions becomes progressively more
important. When collisional damping (computed following
\citealt{pot99}) is accounted for, we found that the surface emissivity
is substantially depressed below $\approx 1$~keV with respect to the cold
electron gas case. At low fields ($\lesssim 5\times 10^{12}$~G) 
virtually no emission
is expected at energies below $\approx 0.5$~keV, while the decline 
at low energies is not so sharp for $B\gtrsim 10^{13}$~G (see Figure~
\ref{alpha_tot} and the discussion at the end of \S\ref{damp}).
The emerging spectrum is shown in Figure~\ref{specdamp} for two
representative values of the polar magnetic field. Despite the
spectra deviating quite strongly from a blackbody distribution at low
energies, the fit with a blackbody in the 0.1-2~keV range is still
acceptable, with maximal deviations typically below 20\%. At the
lower field strength shown in Figure~\ref{specdamp} ($B_p=2\times
10^{13}$~G), $f_E\sim 0.35$ which would imply a radius larger
than the pure blackbody radius 
by about a factor two. This is definitely larger than what is predicted by
the cold electron gas models with $\rho=\rho_s$, and may be enough to 
provide an acceptable value of the stellar radius. However, at least for
the uniform temperature distribution, for such values of the polar field
the absorption feature around $\omega_0\sim 300$~eV is clearly present
in the spectrum (see again Figure~\ref{specdamp}). The feature is 
not so pronounced at larger fields but $f_E$ becomes higher ($\sim
0.45$) making the radius a problem again. One has also to bear in mind 
that the spectra shown in 
Figure~\ref{specdamp} have been computed for a fixed rejection threshold.
As Figure~\ref{rejec} shows, the choice of this parameter (even within a 
factor of a few) has a crucial influence on the shape of emitted 
spectrum.

Apart from the considerable uncertainties in current modeling of the
physics governing the phase transition (see \S \ref{bare} and
\citealt{lai2001} for a more detailed discussion), we remind the reader
that our spectra have been computed under a number of simplifying
assumptions. A thorough discussion of the limitations of this kind of
approach can be found in \citet{bri80}. The greatest uncertainties
arise because of the assumption of a sharp transition from vacuum to a
smooth metallic surface, neglecting the effects of the macroscopic
surface structure. We assumed the surface is made of pure iron, but
different chemical compositions, or the presence of impurities in the
iron surface, may change the results. Inside the star we neglect the
role of bound electrons and further effects produced by the
dissipation of those waves that are rapidly attenuated within a skin
penetration depth.

Finally, from a different perspective and regarding the possible application
of the present work, we point out that the calculation of the complex
refractive indices below the plasma frequency presented here may
substantially contribute to the determination of the photon
thermal conductivities of ultramagnetized neutron stars. These
are, in turn, important for the accurate calculation of the
thermal structure and cooling of these objects (see
\citealt{pot03}).


\begin{acknowledgements}
We are deeply indebted to A. Potekhin for a critical reading of the
manuscript, for his comments and suggestions and for pointing out
the relevance of this work in connection with photon thermal
conductivities computations. We thank M. Chieregato and K. Wu for 
many helpful discussions during the earlier stages of this work. We 
also acknowledge an anonymous
referee whose penetrating questions and comments greatly improved
an earlier version of this paper. This work was partially
supported by the Italian Ministry for Education, University and
Research (MIUR) under grant COFIN-2002-027145. JJD was supported
by NASA contract NAS8-39073 to the {\em Chandra X-ray Center}.
\end{acknowledgements}


\clearpage

{\appendix

\section{Effects of vacuum polarization}
\label{app1}

In \S~\ref{cold} and \S~\ref{damp} the NS surface emissivity has
been computed under the assumption that radiation propagates in
vacuo outside the star and neglecting the magnetized vacuum
birefringence and polarization properties. At the field strengths
we consider ($B\lesssim B_{QED}\equiv m_ec^3/\hbar e\simeq
4.4\times 10^{13}$~G), this has little affect on the vacuum
refractive index for which deviations from unity are very small
(see below). Yet, radiation propagating in the magnetized vacuum
has two well-defined polarization states, corresponding to the
ordinary ($O$) and extraordinary ($X$) modes, even if the two
modes propagate at very nearly the same speed ($n_O\simeq n_X
\simeq 1$). Therefore, in principle the entire formalism should be
generalized to account for non-scalar absorption and emission
coefficients. In order to quantify this effect, we proceed as
follows. We maintain the same cartesian frame introduced in
\S~\ref{cold} and, at fixed magnetic co-latitude $\theta$, we
first consider incident radiation with polarization mode $s$
(here and in the following $s$ stands for either $O$ or $X$). The
vacuum dielectric tensor is expressed as (e.g. \citealt{mes92};
\citealt{hh97})

\begin{equation}
\epsilon^{vac}_{ij} = \left ( \begin{array}{ccc}
a + q \sin^2 \alpha & 0  & q \sin \alpha \cos \alpha \\
0 & a  & 0 \\
q \sin \alpha  \cos \alpha & 0  & a + q \cos^2 \alpha
\end{array}
\right )\, .
\end{equation}
Suitable expressions for $a, q$ have been given by \cite{holai2003}.
In the weak field
limit $B\lesssim B_{QED}$ it is
\begin{equation}
 a\approx 1- 2 \delta_V , \quad q \approx 7
\delta_V , \quad \delta_V = \frac{\alpha_F}{45 \pi} b_V^2
\end{equation}
where $\alpha_F =1/137$ and $b_V = B/B_{QED}$, while for $B\gtrsim
B_{QED}$ it is
\begin{eqnarray}
 a & \approx & 1 +   \frac{\alpha_F}{45 \pi} \left [ 1.195 - \frac{2}{3}
\ln
b_V
- \frac {1}{b_V } \left ( 0.8553 + \ln b_V \right ) - \frac{1}{2 b_V^2}
\right ] \\ \nonumber
q & \approx &  -  \frac{\alpha_F}{45 \pi} \left [- \frac{2}{3} b_V + 1.272
-  \frac {1}{b_V }  \left ( 0.3070 + \ln b_V \right ) - 0.7003 \frac
{1}{b_V^2 } \right ] \,.
\end{eqnarray}
The
two unit propagation eigenmodes are

\begin{eqnarray}
\mathbf{e}_X& =& \frac{\mathbf{k} \times \mathbf{b}}{\sin\delta}
 = \frac{1}{\sin\delta}(\cos\alpha\sin
i\sin\beta\, ,-\cos\alpha\sin i\cos\beta+\sin\alpha\cos i\, ,
-\sin\alpha\sin i\cos\beta) \cr \mathbf{e}_O &=& \frac{\mathbf{b}
-n_O^2\cos\delta \mathbf{k}}{\sqrt{1+ n_O^2(n_O^2-2)\cos^2\delta}}
=\frac{1}{\sqrt{1+n_O^2(n_O^2-2)\cos^2\delta}}
(\sin\alpha-n_O^2\cos\delta\sin i\cos\beta\, ,\cr
&&-n_O^2\cos\delta\sin i\sin\beta\,
,\cos\alpha-n_O^2\cos\delta\cos i)
\end{eqnarray}
with $\cos \delta \equiv {\bf k} \cdot {\bf b} =  \cos i\cos\alpha +\sin i\sin\alpha\cos\beta$.
Series expansions for the refractive indices of the two modes have been
given by \cite{hh97}. In the
weak-field limit it is
\begin{eqnarray}
n_O & \approx & 1 - \frac{\alpha_F} {4 \pi} \sin^2 \delta \left [
\frac{16}{3} B_4 b_V^2 + \frac{64}{5} B_6 b_V^4 + O(b_V^6) \right  ] +
O\left [ \left (  \frac{\alpha_F} {4 \pi} \right )^2 \right ] \\
\nonumber
n_X & \approx& 1 + \frac{ \alpha_F} {4 \pi} \sin^2 \delta \left [
\frac{14}{45} b_V^2  - 0.53 \left (6 B_6-5 B_4 \right)  b_V^4 +
 O(b_V^6) \right ]
+ O\left [ \left (  \frac{\alpha_F} {4 \pi} \right )^2
\right ]
\, ,
\end{eqnarray}
while the corresponding expressions in the strong-field limit are

\begin{eqnarray}
n_O & \approx& 1 + \frac{ \alpha_F} {4 \pi} \sin^2 \delta \left [
\frac{2}{3} - \left ( \ln  b_V + 1 - \ln \pi \right) \frac{1}{b_V}
+ O \left (  \frac{1}{b_V^2} \right )
\right
]
+ O\left [ \left (  \frac{\alpha_F} {4 \pi} \right )^2
\right ]
\\ \nonumber
n_X & \approx&  1 + \frac{ \alpha_F} {4 \pi} \sin^2 \delta \left [
\frac{2}{3} b_V  - \left (8 \ln A - \frac{1}{3} - \frac{2}{3}
\gamma_E \right ) - \left ( \ln \pi + \frac{\pi^2}{18} - 2 - \ln
b_V \right ) \frac{1}{b_V}\right.\cr && \left. + O \left (
\frac{1}{b_V^2} \right ) \right ] + O\left [ \left (
\frac{\alpha_F} {4 \pi} \right )^2 \right ] \, ;
\end{eqnarray}
here $B_n$ are the Bernoulli numbers, $\ln A \simeq 0.2488$ and
$\gamma_E$ is the Euler constant.

At the surface an incident electromagnetic wave, described by its
electric field $\mathbf{E}_s$, wave vector $\mathbf{k}_s$ and
refraction index $n_s$, is partly reflected and partly refracted.
Because of the birefringence of both media (the vacuum and the
solid star crust) this gives rise to two refracted and two
reflected waves. The latter are again either $X$- or
$O$-polarized with electric field $\mathbf{E}_{s,r}''$ where,
once more, $r=O,\, X$\footnote{Strictly speaking, the reflected wave
is not necessary linearly polarized; however it will separate in
the two allowed polarization states after propagating a distance
$l_v\approx 2 \pi c/(\omega | n_O - n_X|)\approx 10^{-4}$ cm for
X-ray energies and $B\approx B_{QED}$ (see e.g. \citealt{chan79};
\citealt{mes92}). The total intensity of the reflected radiation
will be the same far away from the source, despite the change in
the polarization state.}. In the following we ignore the small
($O\vert n_O-n_X\vert$) difference in the angle of reflection,
and assume that the two reflected waves propagate in the same
direction fixed by the angles $i$, $\beta +\pi$.

Since we are not interested in studying the polarization state of the
emergent radiation, but only its spectral distribution,
it is useful to introduce the two quantities
\begin{equation}
\rho_{\omega,s} \equiv \displaystyle{\frac
{\sum_r|\mathbf{E}_{s,r}''|^2}{|\mathbf{E}_s|^2}}
\end{equation}
which represent the ratios between the reflected and incident
intensities when the incident wave is either $X$- or
$O$-polarized (intensities are proportional to squared amplitudes
of the electric fields to first approximation \footnote{We
neglect deviations between the directions of the two
reflected waves and that of the corresponding time-averaged energy fluxes,
the latter defined by the Poynting vectors.}). The absorption
coefficients  corresponding to the two incident modes are
$\alpha_{\omega,s}=1-\rho_{\omega,s}$ and, again, from
Kirchhoff's law we get the total emissivity $j_\omega =
\alpha_\omega B_\omega(T)$, where $\alpha_\omega =
(\alpha_{\omega,O} + \alpha_{\omega,X})/2$. The monochromatic and
total flux are computed again by performing the integrals in
eqs.~(\ref{df}) and (\ref{ftot}).

Inside the star, each refracted wave splits into an ordinary ($
\mathbf{E}'_{s,1}, \mathbf{k}'_{s,1}$) and an extraordinary
($\mathbf{E}'_{s,2}, \mathbf{k}'_{s,2}$) mode. In order to
compute $\rho_{\omega,s}$, we then proceed exactly as discussed
in \S~\ref{cold} by solving the dispersion relation and computing
the refractive index $n_i$, $i=1,2$, for the two refracted modes.
Since $n_s$ now explicitly appears in Snell's law ($n = n_s \sin
i /\sin \Theta$), we have to solve the dispersion relation twice
\begin{eqnarray}\label{dispvac}
& n^4(P+v\sin^2\alpha)+n^2(gv-2PS+u\sin^2\alpha)+PRL+gu =\cr &n_s
\sin i\sin(2\alpha)\cos\beta(n^2-n_s^2 \sin^2i)^{1/2}(u+n^2v)\, ,
\end{eqnarray}
where $g=n_s^2 \sin^2i[1-\sin^2\alpha(1+\cos^2\beta)]$ and all other
quantities are the same as in eq.~(\ref{disp}).

Once the refractive indices are known,
we solve the wave equations $\lambda_{s,ij}(n_m)E'_{s,mj}=0$, where
$E'_{s,mj}$ are the cartesian components of ${\bf E}'_{s,m}$, obtaining
the two ratios $E'_{s,mx}/E'_{s,mz}$ and $E'_{s,my}/E'_{s,mz}$. The
resulting expressions are
\begin{eqnarray}\label{aandbvac}
\frac{E'_{s,mx}}{E'_{s,mz}} \equiv a_m & = &
 \left[ - n_m^2  n_s^2 \sin^2 i \sin \beta \cos \beta -i D  n_s^2 \sin^2 i
\cos \alpha
+ i D n_s \cos \beta \sin \alpha  \sin i \sqrt{n_m^2 -  n_s^2 \sin^2
i}
\right.\nonumber\\
& - & \left. n_s \sin
\beta \left (P-S \right ) \sin \alpha \cos \alpha   \sin i
\sqrt{n_m^2 -
 n_s^2 \sin^2 i}\right.\nonumber\\
&+& \left.  n_s^2 \sin^2 i \sin \beta \cos \beta \left (P \cos^2 \alpha +
S \sin^2
\alpha \right ) + i D \cos \alpha P\right ]\nonumber\\
& \times &\left [- n_m^2  n_s \sin i \sqrt{n_m^2 -  n_s^2 \sin^2 i} \sin
\beta + i D
\sin
\alpha n_m^2 - iD  n_s^2 \sin \alpha \sin^2 i \cos^2 \beta\right.
\nonumber\\
&- & \left.i D n_s \cos \alpha \cos \beta  \sin i \sqrt{n_m^2 -
n_s^2 \sin^2
i}
+ n_s \sin i\sqrt{n_m^2 -  n_s^2 \sin^2 i} \left [ \sin \beta S
\right.\right.\nonumber\\
&+& \left.\left.
\sin \beta \sin^2 \alpha\left (P-S \right )\right] -\left(P-S \right)
n_s^2
\sin \alpha \cos \alpha  \sin^2 i \sin \beta \cos
\beta -iD \sin \alpha P\right]^{-1} \\
\frac{E'_{m,y}}{E'_{m,z}}  \equiv  b_m  & = &  \left[a_m\left(
n_s^2 \sin^2i
\sin\beta\cos\beta-iD\cos\alpha\right)+n_s \sin\beta  \sin
i\sqrt{n_m^2- n_s^2 \sin^2 i}+
iD\sin\alpha\right]\nonumber\\
&\times & \left( n_s^2 \sin^2\beta\sin^2 i-n_m^2+S\right)^{-1}\, .
\end{eqnarray}

The generalization of the Fresnel equations (e.g. \citealt{ja75}
and B80) to the present case gives

\begin{eqnarray}\label{fresnel}
n_s E_{s\perp} & + & \sum_r n_r E_{s,r\perp}^{''} = n_s \sum_m \bc_m
E'_{s,mz} \nonumber\\
n_s E_{s\perp} & - & \sum_r n_r E_{s,r\perp}^{''} = \sum_m w_m \bc_m
E'_{s,mz} \nonumber\\
E_{s\parallel} & - & \sum_r E_{s,r\parallel}^{''} =
\sum_m \frac{\ac_m}{\cos i}E'_{s,mz} \nonumber \\
E_{s\parallel} & + & \sum_r E_{s,r\parallel}^{''}  = \sum_m
\frac{\cc_m}{\sin i} E'_{s,mz}\, .
\end{eqnarray}
In the previous expressions the components of $\mathbf{E}_s$,
$\mathbf{E}''_{s,r}$ are parallel and orthogonal to the plane of
incidence, $w_m=\sqrt{n_m^2 - n_s^2 \sin^2 i}/\cos i$, $\ac_m =
b_m\sin\beta-a_m\cos\beta$, $\bc_m = b_m\cos\beta+a_m\sin\beta$,
$\cc_m = a_m \epsilon'_{31} + b_m \epsilon'_{32} + \epsilon'_{33}$
and $\epsilon'_{ik}\epsilon^{vac}_{kj}=\epsilon_{ij}$. The
components of the electric field of the incident and reflected waves can be
expressed in terms of the amplitudes $E_s$ and $E_{s,r}''$ as
\begin{eqnarray}\label{pareperp}
E_{s\perp} = f_s E_s, &\quad& E_{s\parallel} = g_sE_s\cr
E_{s,r\perp}= f_r'' E_{s,r}'',&\quad& E_{s,r\parallel} = g_r''E_{s,r}''
\end{eqnarray}
where
\begin{eqnarray}\label{projec}
f_X = \cos\gamma, &\quad& g_X=\sin\gamma, \quad \cos\gamma=
\frac{-\cos\alpha\sin i+\cos\beta\sin\alpha\cos i}{\sin\delta}\cr
f_O = \cos\xi, &\quad& g_O=\sin\xi, \quad \cos\xi=
\frac{\sin\beta\sin\alpha}{\sqrt{1+
n_O^2(n_O^2-2)\cos^2\delta}}\, .
\end{eqnarray}
The expressions for $f_X''$, $f_O''$, $g_X''$, $g_O''$ are obtained
from the previous ones by replacing $\beta$ with $\beta+\pi$.

Inserting (\ref{pareperp}) into equations (\ref{fresnel}) finally
gives the amplitude ratios

\begin{eqnarray}
\label{oxrefl}
\frac{E_{s,O}''}{E_s} &=& - \frac{
 \left [
f_s \displaystyle{\frac{ \overline w_2 -1 }
{ (\overline w_2 - \overline w_1)  \bc_1}} -
g_s   \frac{\cc_2 \cos i - \ac_2 \sin i}{\ac_1 \cc_2 - \ac_2 \cc_1} \right
] \left
(\overline Q_s -
Q_X \right ) }
{
\left [\displaystyle{\frac{n_O}{n_s}} f_O^{''}  \frac{
\overline w_2 + 1 }{ (\overline w_2 - \overline w_1 )\bc_1} +
g_O^{''}    \frac{\cc_2 \cos i + \ac_2 \sin i}{\ac_1 \cc_2 - \ac_2
\cc_1}   \right ] \left( Q_O - Q_X \right ) }\cr
\frac{E_{s,X}''}{E_s}& = &-\frac{
 \left [
f_s \displaystyle{\frac{ \overline w_2 -1 }{
(\overline w_2 - \overline w_1)  \bc_1}} -
g_s   \frac{\cc_2 \cos i - \ac_2 \sin i}{\ac_1 \cc_2 - \ac_2 \cc_1} \right
] \left (
\overline Q_s -
Q_O \right ) }
{
\left [\displaystyle{
\frac{n_X}{n_s}} f_X^{''}   \frac{\overline w_2 +
1 }{ (\overline w_2 - \overline w_1) \bc_1} +
 g_X^{''}   \frac{\cc_2\cos i + \ac_2 \sin i}{\ac_1 \cc_2 - \ac_2
\cc_1} \right ]  \left( Q_X - Q_O \right ) }
\end{eqnarray}
where $\overline w_m=w_m/n_s$ and
\begin{eqnarray}
\label{qxqo}
Q_X &=& \left [\frac{n_X}{n_s} f_X^{''}   \frac{ \overline w_1 +
1 }{(\overline w_1 - \overline w_2 ) \bc_2} +
g_X^{''}   \frac{\cc_1 \cos i + \ac_1 \sin i}{\ac_2 \cc_1 - \ac_1 \cc_2}
\right ]
\left [\frac{n_X}{n_s} f_X^{''}   \frac{\overline w_2 +
1 }{ (\overline w_2 - \overline w_1) \bc_1} +
g_X^{''}   \frac{\cc_2\cos i + \ac_2 \sin i}{\ac_1 \cc_2 - \ac_2
\cc_1} \right ]^{-1}\cr
Q_O &= &\left [\frac{n_O}{n_s} f_O^{''}  \frac{
\overline w_1 + 1}{(\overline w_1 - \overline w_2 ) \bc_2}
+ g_O^{''}    \frac{\cc_1 \cos i + \ac_1 \sin i}{\ac_2 \cc_1 - \ac_1
\cc_2} \right ]
\left [ \frac{n_O}{n_s} f_O^{''}  \frac{
\overline w_2 + 1 }{ (\overline w_2 - \overline w_1 )\bc_1} +
g_O^{''}    \frac{\cc_2 \cos i + \ac_2 \sin i}{\ac_1 \cc_2 - \ac_2
\cc_1}\right ]^{-1} \cr
\overline Q_s &= &
\left [ f_s \displaystyle{\frac{
\overline w_1 -1}{(\overline w_1 -
\overline w_2 ) \bc_2}}
- g_s  \frac{\cc_1 \cos i - \ac_1 \sin i}{\ac_2 \cc_1 - \ac_1 \cc_2}
\right ]
 \left [
f_s \displaystyle{\frac{ \overline w_2 -1 }
{ (\overline w_2 - \overline w_1)  \bc_1}} -
g_s   \frac{\cc_2 \cos i - \ac_2 \sin i}{\ac_1 \cc_2 - \ac_2 \cc_1} \right
]^{-1}
\, .
\end{eqnarray}

Again, when only one
refracted wave (labeled ``1'' for convenience) survives,
the previous calculation yields
($f_s^{''}, g_s^{''} \neq 0$)

\begin{eqnarray}
\label{onemodevac}
\frac{E_{s,O}''}{E_s} &=&
=  \frac{\displaystyle{
\frac{n_s}{n_X}} \frac{f_s}{f^{''}_X}  \frac{
\overline w_1  -1 }{ \overline w_1  + 1 }+
\frac{g_s}{ g_X^{''}}   \frac{ \cc_1 \cos i - \ac_1 \sin i }{ \cc_1 \cos i
+
\ac_1 \sin i}}
{\displaystyle{\frac{ g_O^{''}}{ g_X^{''}}}-
 \frac{n_O}{n_X}  \frac{f_O^{''}}{f^{''}_X}}\cr
\frac{E_{s,X}''}{E_s } & =&\frac{\displaystyle{\frac{n_s}{n_O}}
\frac{f_s}{f^{''}_O}  \frac{
\overline w_1  -1 }{ \overline w_1  + 1 }  + \frac{g_s}{ g_O^{''}}
\frac{ \cc_1 \cos i - \ac_1 \sin i }{ \cc_1 \cos i +
\ac_1 \sin i}}{\displaystyle{\frac{ g_X^{''}}{ g_O^{''}}} -
\frac{n_X}{n_O}
\frac{f_X^{''}}{f^{''}_O}}
\end{eqnarray}
from which $\rho_{\omega,s}$ follows. Similar expressions can be derived
in the case in which either $f_s^{''}$ or $g_s^{''}$ vanishes.

The absorption coefficients $\alpha_{\omega,s}=1-\rho_{\omega,s}$
have been computed numerically in the relevant angular ranges
following the procedure outlined above and the results have been
used to evaluate $j_\omega$ and $f_\omega$ (see eq. [\ref{df}]).
The quantity $f_\omega/B_\omega(T)$ has then been compared with
its value computed by neglecting vacuum polarization. We repeated
the comparison for different values of magnetic field; some
examples are shown in Figure~\ref{vacuum}. Fractional corrections
turn out to be always negligible, being at most $\sim 10^{-2}$
below 2 keV. Vacuum corrections enter the dielectric tensor via
the two quantities $a-1$ and $q$, and, even for $B_p \sim 5
\times 10^{13}$~G, the largest value we consider in our model, it
is $\vert a-1\vert \approx q\approx 10^{-4}$. The corresponding
deviation of the refractive index from unity in the vacuum
outside the star is indeed negligible. Therefore, when we compute
the total reflectivity we are superimposing two incident modes
perpendicular to each other. By choosing them in the plane
perpendicular and parallel to the plane of incidence or, as in
this case, with ordinary and extraordinary polarization, is
not important as long as they travel at nearly the same speed. }

\clearpage

\begin{deluxetable}{lccl}
\tablecolumns{4}
\tablewidth{0pc}
\tablecaption{Isolated Neutron Stars Parameters\label{tableins}}
\tablehead{
\colhead{Source} &
\colhead{$T_{bb}$ (eV)\tablenotemark{a}} &
\colhead{$B \, (10^{12}$ G)\tablenotemark{b}} &
\colhead{Refs.\tablenotemark{c}}
}
\startdata
         \oneight   & $61.1\pm 0.3$ &      --       & 1, 2 \\
         \zeroseven & $86.0\pm 0.6$ & $21.3 \pm 0.1 $ & 3, 4 \\
         Vela       & $128.4\pm 7 $ & $3.3   $ & 5, 6    \\
         Geminga    & $48.3^{+6.1}_{-9.5}$ & $1.5$ & 7, 8 \\
         PSR 0656+14    & $69.0\pm 2.5$ & $4.7$ & 9, 6     \\
         PSR 1055-52    & $68.1^{+10.2}_{-17.2}$ & $ 1.1 $ & 10, 6  \\
\enddata
\tablenotetext{a}{Errors refer to $2\sigma$ confidence level}
\tablenotetext{b}{As computed from the spin-down formula; period
derivative is very accurate for the three radiopulsars and for Geminga.
Errors refer to 90\% confidence level for \zeroseven.
}
\tablenotetext{c}{
[1] \citet{bur2001}; [2] \citet{dra2002}; [3] \citet{pae2001};
[4] \citet{silvia2002}; [5] \citet{pa2001};
[6] \citet{pulcat93}; [7] \citet{hw97}; [8] \citet{bicar96};
[9] \citet{masch2002}; [10] \citet{gr96}
}
\end{deluxetable}

\clearpage


\begin{figure}
\includegraphics[width=6in,angle=0]{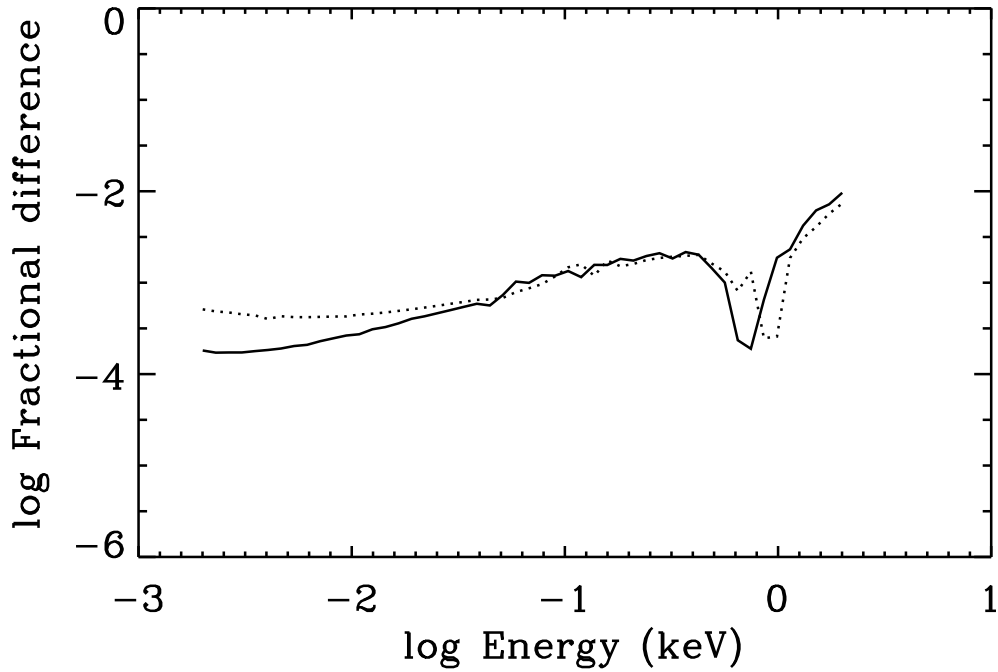}
\caption{\label{vacuum} Fractional difference in the computed values of
$F_\omega/B_\omega(T)$  with and without vacuum polarization taken
into account (see
Appendix \ref{app1}) at different energies. Solid
and dotted lines are for  $B_p = 3 \ {\rm and} \ 6 \times
10^{13}$~G respectively.
}
\end{figure}

\end{document}